\documentclass[prd,a4paper,aps,showpacs,floats,superscriptaddress,nofootinbib]{revtex4}
\usepackage{epsfig}
\usepackage{placeins}
\usepackage{amssymb}
\usepackage{color}
\usepackage{float}
\usepackage{hyperref}
\usepackage{yfonts}
\usepackage{amsmath}
\usepackage{MnSymbol}
\usepackage{amssymb}
\usepackage{amsfonts}
\usepackage{eufrak}
\usepackage{bm}
\usepackage{phoenician}
\usepackage{graphicx}

\newcommand{\bV}{{\mathbf V}}

\newcommand{\bn}{{\mathbf n}}

\newcommand{\bs}{\mathbf{s}}
\newcommand{\br}{\mathbf{r}}
\newcommand{\ud}{\mathrm{d}}

\newcommand{\aaa}{$\alpha$ }

\newcommand{\de}{\delta}
\newcommand{\De}{\Delta}

\newcommand{\be}{\begin{equation}}
\newcommand{\ee}{\end{equation}}

\newcommand{\bea}{\begin{eqnarray}}
\newcommand{\eea}{\end{eqnarray}}
\newcommand{\bean}{\begin{eqnarray*}}
\newcommand{\eean}{\end{eqnarray*}}
\newcommand{\dd}{\partial}

\newcommand{\HH}{{\cal H}}

\usepackage{calc}
\newlength{\depthofsumsign}
\newcommand{\nsum}[1][1.4]{
    \mathop{%
        \raisebox
            {-#1\depthofsumsign+1\depthofsumsign}
            {\scalebox
                {#1}
                {$\displaystyle\sum$}%
            }
    }
}

\makeatletter
\newcommand*{\DivideLengths}[2]{%
  \strip@pt\dimexpr\number\numexpr\number\dimexpr#1\relax*65536/\number\dimexpr#2\relax\relax sp\relax
}
\makeatother

\usepackage{color}

\definecolor{darkred}{RGB}{175,0,0}
\definecolor{darkblue}{RGB}{14,0,185}

\begin{document}

\title{Doppler term in the galaxy two-point correlation function: wide-angle, velocity, Doppler lensing and cosmic acceleration effects}

\author{Alvise Raccanelli}
\affiliation{Department of Physics \& Astronomy, Johns Hopkins University, Baltimore, MD 21218, USA}

\author{Daniele Bertacca}
\affiliation{Argelander-I  b 22nstitut f\"ur Astronomie, Auf dem H\"ugel 71, D-53121 Bonn, Germany}

\author{Donghui Jeong}
\affiliation{Department of Astronomy and Astrophysics, The Pennsylvania State University, University Park, PA 16802, USA}
\affiliation{Institute for Gravitation and the Cosmos, The Pennsylvania State University, University Park, PA 16802, USA}

\author{Mark C. Neyrinck}
\affiliation{Department of Physics \& Astronomy, Johns Hopkins University, Baltimore, MD 21218, USA}

\author{Alexander S. Szalay}
\affiliation{Department of Physics \& Astronomy, Johns Hopkins University, Baltimore, MD 21218, USA}

\date{\today}

\begin{abstract}
We study the parity-odd part (that we shall call {\it Doppler term})
of the {\it linear} galaxy two-point correlation function that arises from 
wide-angle, velocity, Doppler lensing and cosmic acceleration effects.
As it is important at low redshift and at large angular separations, the 
Doppler term is usually neglected in the current generation of galaxy surveys. 
For future wide-angle galaxy surveys such as Euclid, SPHEREx and SKA, however,
we show that the Doppler term must be included.
The effect of these terms is dominated by the magnification due to relativistic aberration
effects and the slope of the galaxy redshift distribution and it generally mimics 
the effect of the local type primordial non-Gaussianity with the 
effective nonlinearity parameter $f_{\rm NL}^{\rm eff}$ of a few; we show that this would affect forecasts on measurements of $f_{\rm NL}$ at low-redshift.
Our results show that a survey at low redshift with large number density over 
a wide area of the sky could detect the Doppler term with a 
signal-to-noise ratio  of $\sim 1-20$, depending on survey specifications.
\end{abstract}


\maketitle

\section{Introduction}
Measuring anisotropies in galaxy clustering due to the Redshift-Space 
Distortions (RSD) is one of the prime scientific goals of current and future galaxy surveys, and the galaxy correlation function is by now one of the most important probes used to test cosmological models (see e.g.~\cite{Peacock:2001, Percival:2dF, Tegmark:2006, Guzzo:2008, Blake:2011a, Blake:2011b, Blake:2011c, Samushia:2012, Reid:2012, Jennings:2012, Zhao:2012, dePutter:2012, Sanchez:2013, Blake:2013, Anderson:2013, Samushia:2013, Raccanelli:growth, Cyr-Racine:2014, Dvorkin:2014, Samushia:2014, Tojeiro:2014, Cuesta:2015, Gil-Marin:2015}).

Given that forthcoming galaxy surveys aim to measure galaxy clustering with high precision~\cite{euclid, Raccanelli:SKA, pfs, Font-Ribera:2013}, a very precise theoretical modeling will be required in order to interpret the results correctly.
The standard way to model the 2-point correlation function at large scales was defined in~\cite{Kaiser:1987, Hamilton:1992, Hamilton:1997}; this modeling was developed for past surveys that were far narrower and shallower than the wide and deep surveys currently being built and planned. For this reason, a series of approximations were adopted, including e.g. using the distant observer (flat-sky) approximation, and limiting to scales large enough to be linear, and small enough not to require a wide-angle and general relativistic treatment.
Going beyond the linear terms is required when trying to model the quasi-linear regime, and several approaches have been suggested in the past (see e.g.~\cite{Neyrinck:2009, Taruya:2010, Jennings:2010, Kwan:2011, Wang:2011, Reid:2011, Neyrinck:2011, Wang:2012, McCullagh:2012, Carlson:2012, Bianchi:2012, delaTorre:2012, Gil-Marin:2012, Wang:2013, Reid:2014, Okumura:2015, Vlah:2015, Bianchi:2014, McCullagh:2014, Jeong:2014, Senatore:2014, Angulo:2015}).
On larger scales, the flat-sky approximation has been dropped and models for the so-called wide-angle RSD, including geometry effects, have been developed and tested~\cite{Szalay:1997, Matsubara:1999, Bharadwaj:1999, Szapudi:2004, Papai:2008, Raccanelli:2010, Montanari:2012, Samushia:2012, Yoo:2013, Raccanelli:3D, Bonvin:2014, Samushia:2015, Reimberg:2015, Slepian:2015}; more recently, formalisms including large-scale relativistic effects have been studied~\cite{Yoo:2009, Yoo:2010, Bonvin:2011, Challinor:2011, Yoo:2012, Jeong:2012, Bertacca:2012, Raccanelli:radial, DiDio:2014, Camera:2015, Raccanelli:2015GR, Alonso:2015, Fonseca:2015, Baker:2015, Bull:2015}.

It is by now clear that future measurements will need to include a series of corrections to the Kaiser formalism on a variety of scales: recent works showed how even small correction due to 
the relative motion between baryon and dark matter at early (post-recombination) times can be relevant for measurements of the BAO peak~\cite{Blazek:2015}. 
While peculiar velocities modify the apparent radial position of a galaxy, Doppler lensing effects modify its apparent size and luminosity due to the relative motion between the galaxy and the observer.
Terms proportional to radial peculiar velocities ($v_r/r$) in the galaxy correlation function are also neglected in the Kaiser formalism and have been proven small for the SDSS-II survey~\cite{Samushia:2012}.

In this paper we write the wide-angle two-point galaxy correlation function in a way that clearly separates the Doppler terms from the density and ``standard" RSD, and investigate their relative importance for future wide galaxy redshift surveys.
The paper is organized as follows: in Section~\ref{sec:fs_wa} we introduce the Doppler term to the galaxy correlation function in the flat-sky and wide angle cases, and in Section~\ref{sec:xi_wa} we present a formalism to distinguish between the Doppler term and the standard RSD expression. Section~\ref{sec:alpha} describes the various effects in more detail, with their detectability being described in Section~\ref{sec:snr}.
Finally, in Section~\ref{sec:conclusions} we draw our conclusions and discuss results and future prospects.


\section{Geometry and velocity terms in flat sky and wide-angle}
\label{sec:fs_wa}
The standard theoretical model for the large-scale 
Redshift-Space Distortion (RSD) analysis is the
so-called Kaiser formula~\cite{Kaiser:1987, Hamilton:1997}, 
that defines the RSD operator relating the real- to the redshift- space 
overdensity, as:
\begin{equation}
\label{eq:kaiser_operator}
{\bf S}^{\rm Kaiser} = 1 + \beta (\hat{\bf r}\cdot\nabla)^2\nabla^{-2} \, ,
\end{equation}
where $\beta = f/b$, $b$ being the galaxy bias and $f = d \, \ln D / d \, \ln a$ is the logarithmic derivative of the growth factor; 
$\hat{\bf r}$ is the line-of-sight (or radial) directional unit vector.

Equation~\eqref{eq:kaiser_operator} can be derived starting from the conservation of the number of galaxies (RSD change the apparent radial position of galaxies, they don't create nor destroy them):
\begin{equation} \label{eq:nd}
N^{\mathcal{S}}(\bs) d^3 s = N^{\mathcal{R}}(\br) d^3 r \, ,
\end{equation}
where $N^{\mathcal{S, \mathcal{R}}}$ are the galaxy density in redshift- and 
real- space, respectively.
The relation between real-space ($\mathcal{R}$) and redshift-space 
($\mathcal{S}$) position is given by
\begin{equation}
  \bs(\br) = \br + v_r(\br) \hat{\br} \, ,
\label{eq:srvr}
\end{equation}
where $v_r\equiv\hat{\bf r}\cdot{\bf v}/(aH)$ is the radial component of the peculiar velocity normalized with the 
Hubble parameter.
Combining Equations~\eqref{eq:nd}-\eqref{eq:srvr}, we obtain:
\begin{align}\label{eq:J0}
  1+\delta^{\mathcal{S}}(\bs) = [1+\delta^{\mathcal{R}}(\br)] 
    \left(1+ \frac{\partial v_r}{\partial r} \right)^{-1} \left(1+ \frac{v_r}{r} \right)^{-2} \frac{\bar{N}(r)}{\bar{N}(r+v_r)} \, ,
\end{align}
where ${\bar{N}}$ is the average number density, which gives, to linear order:
\begin{equation}
\label{eq:J}
  \delta^{\mathcal{S}}(\br) = \delta^{\mathcal{R}}(\br)-
    \left( \frac{\partial v_r}{\partial r}+\frac{\alpha(\br)v_r}{r} \right) \, ,
\end{equation}
with:
\begin{equation}
\label{eq:alphaN}
  \alpha(\br) = \frac{\partial \ln r^2 \bar{N}^s(\br)}{\partial \ln r} \, .
\end{equation}
Usually the $(1+v_r/r)^{-2}$ term in Equation~\eqref{eq:J0}
is omitted because, in the linear regime, it gives rise to a term proportional 
to $v_r/r$, that would tend to be small at large distances for surveys that 
probe separations far smaller than the radial distance to the survey 
observed volume.
However, for future wide surveys probing wide angular scales, the $v_r/r$ term can be of the same order as the $\partial_r v_r$ term, and in general 
cannot be neglected~\cite{Papai:2008, Raccanelli:2010}.

It is important to note that, even when assuming the flat-sky approximation, 
the \aaa term in Equation~(\ref{eq:J}), that we called Doppler term throughout
this paper, 
does not naturally disappear.
This term, however, is indeed small when the flat-sky approximation is 
valid; thus, it is usually neglected.
On the other hand, this term has been explicitly included in wide-angle 
formalisms and it was proven to be a necessary ingredient to properly fit the correlation function from simulations, 
where the model is tested to the high precision required by future large-scale galaxy surveys~\cite{Papai:2008, Raccanelli:2010}. It was however shown that for past galaxy catalogs, the impact of these terms was subdominant with respect to statistical errors~\cite{Samushia:2012, Beutler:2013}.

\begin{center}
\begin{figure*}[htb!]
\includegraphics[width=0.49\columnwidth]{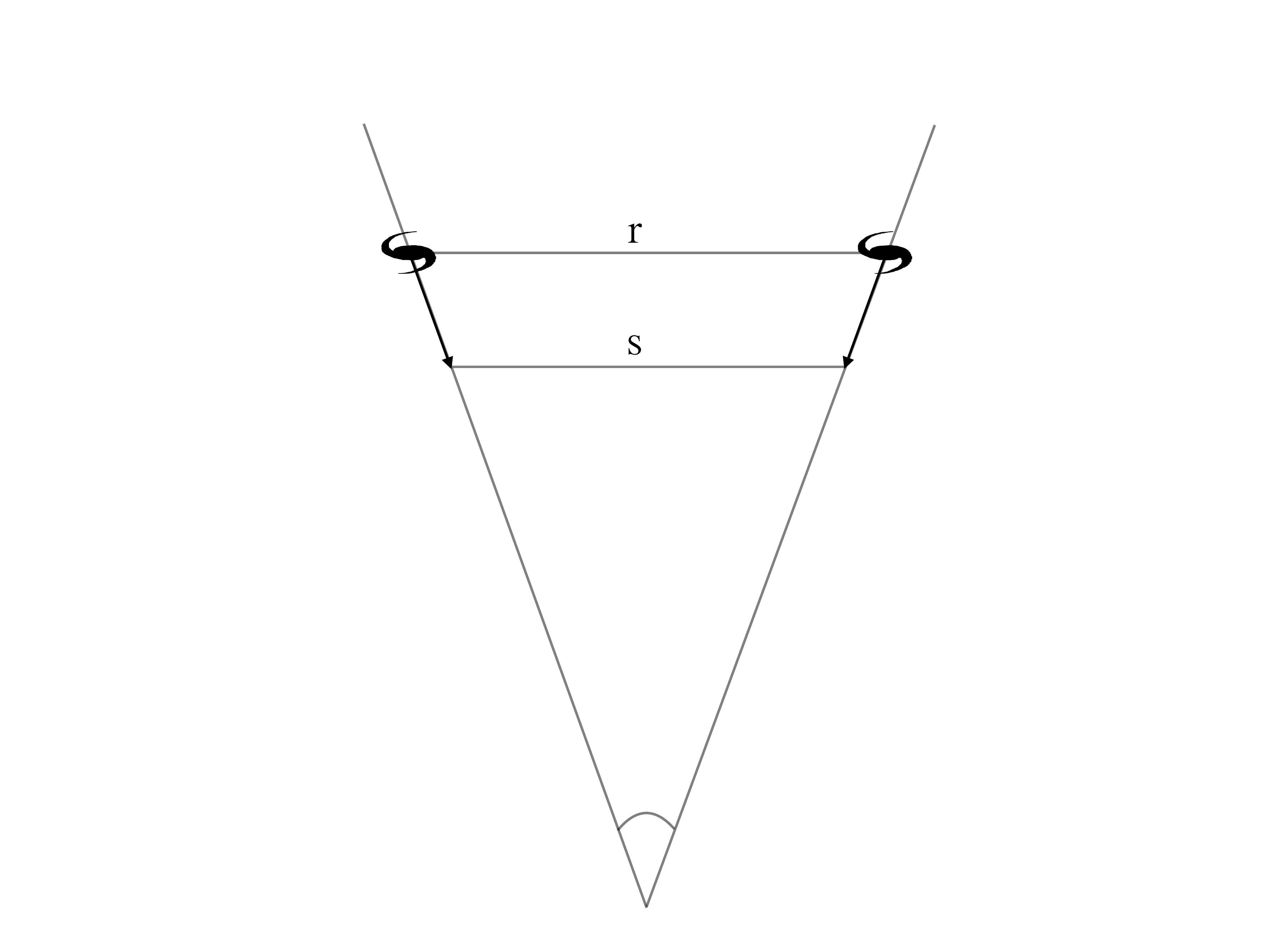}
\includegraphics[width=0.49\columnwidth]{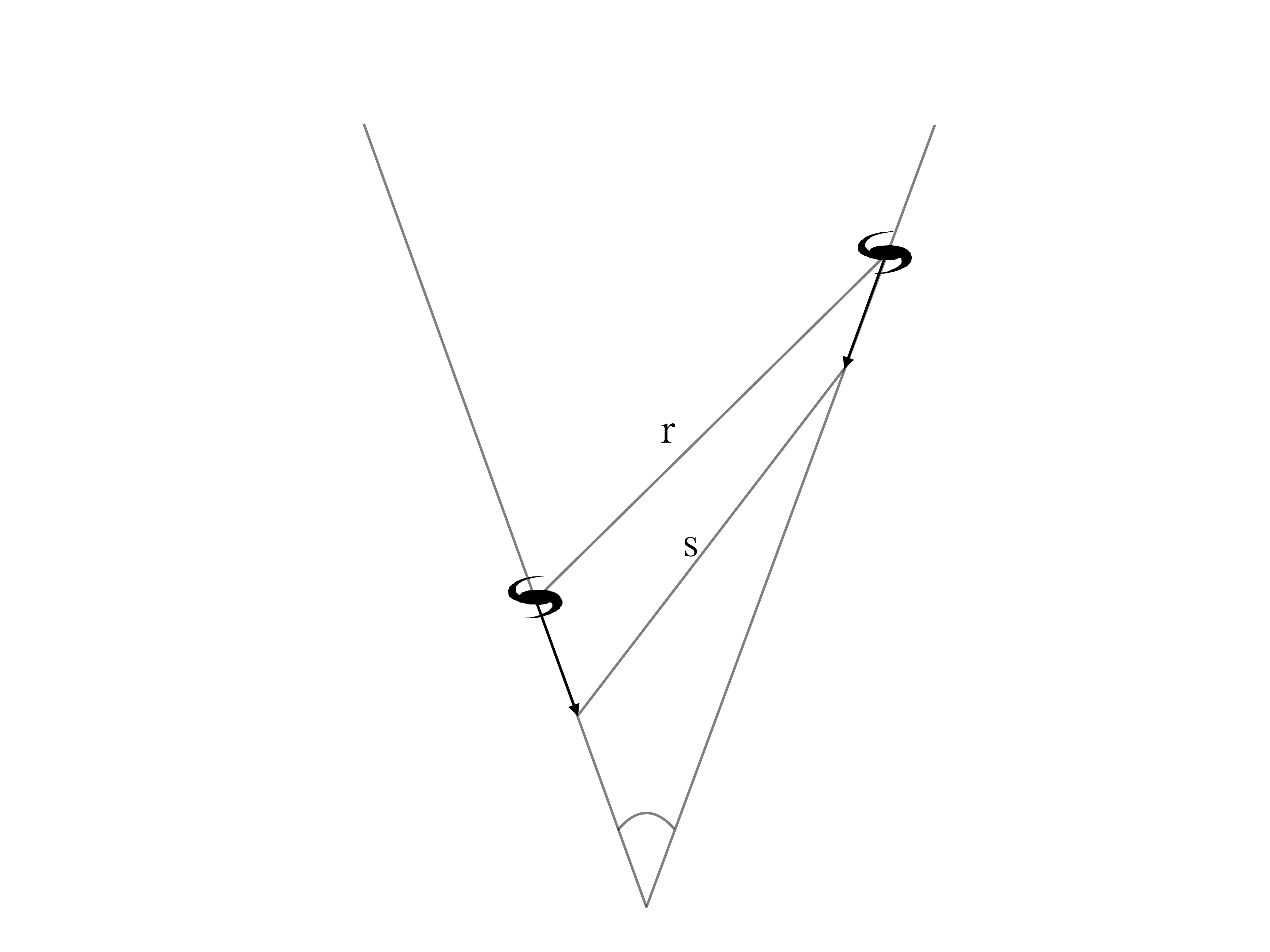}
\includegraphics[width=0.49\columnwidth]{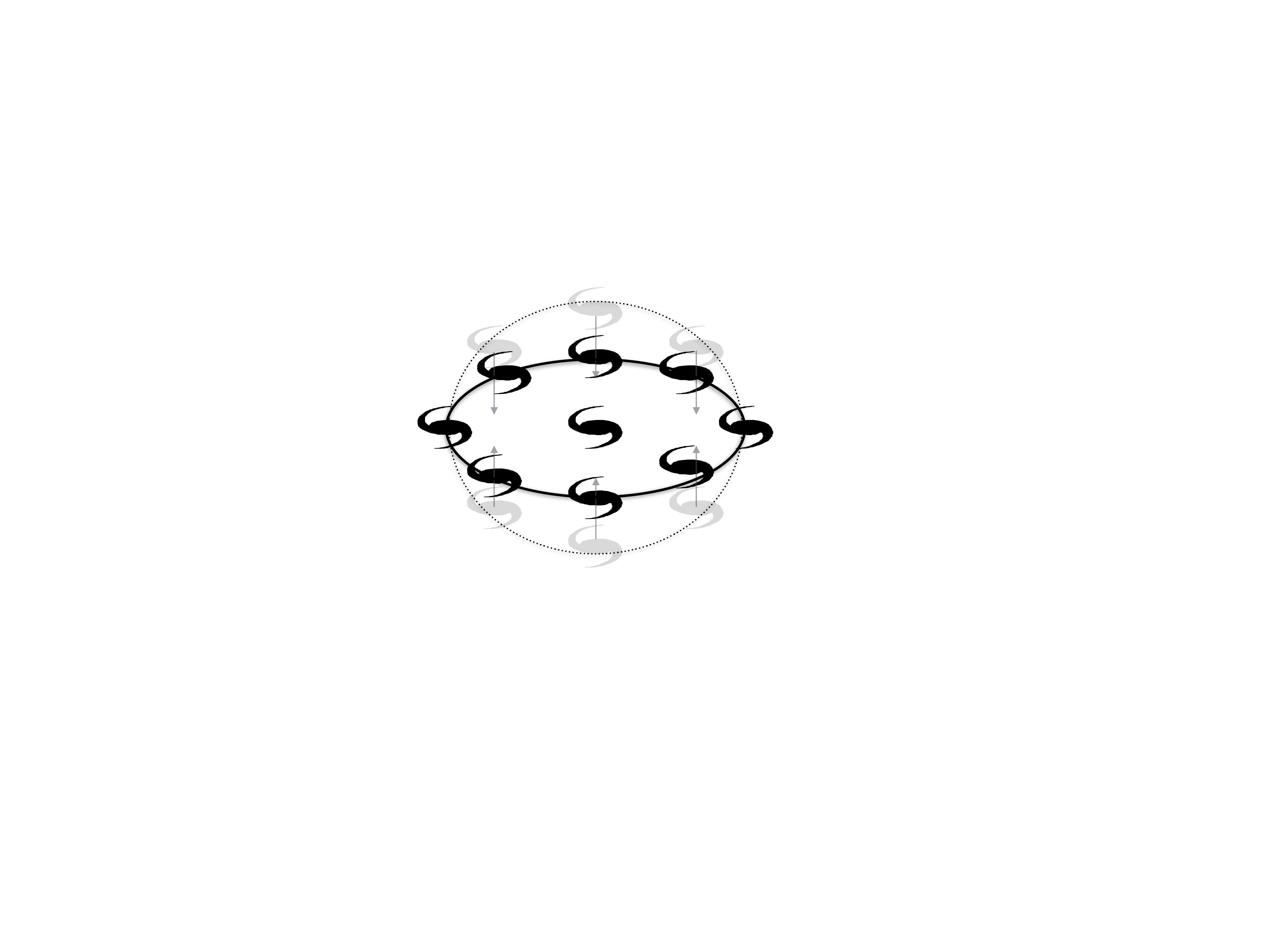}
\includegraphics[width=0.49\columnwidth]{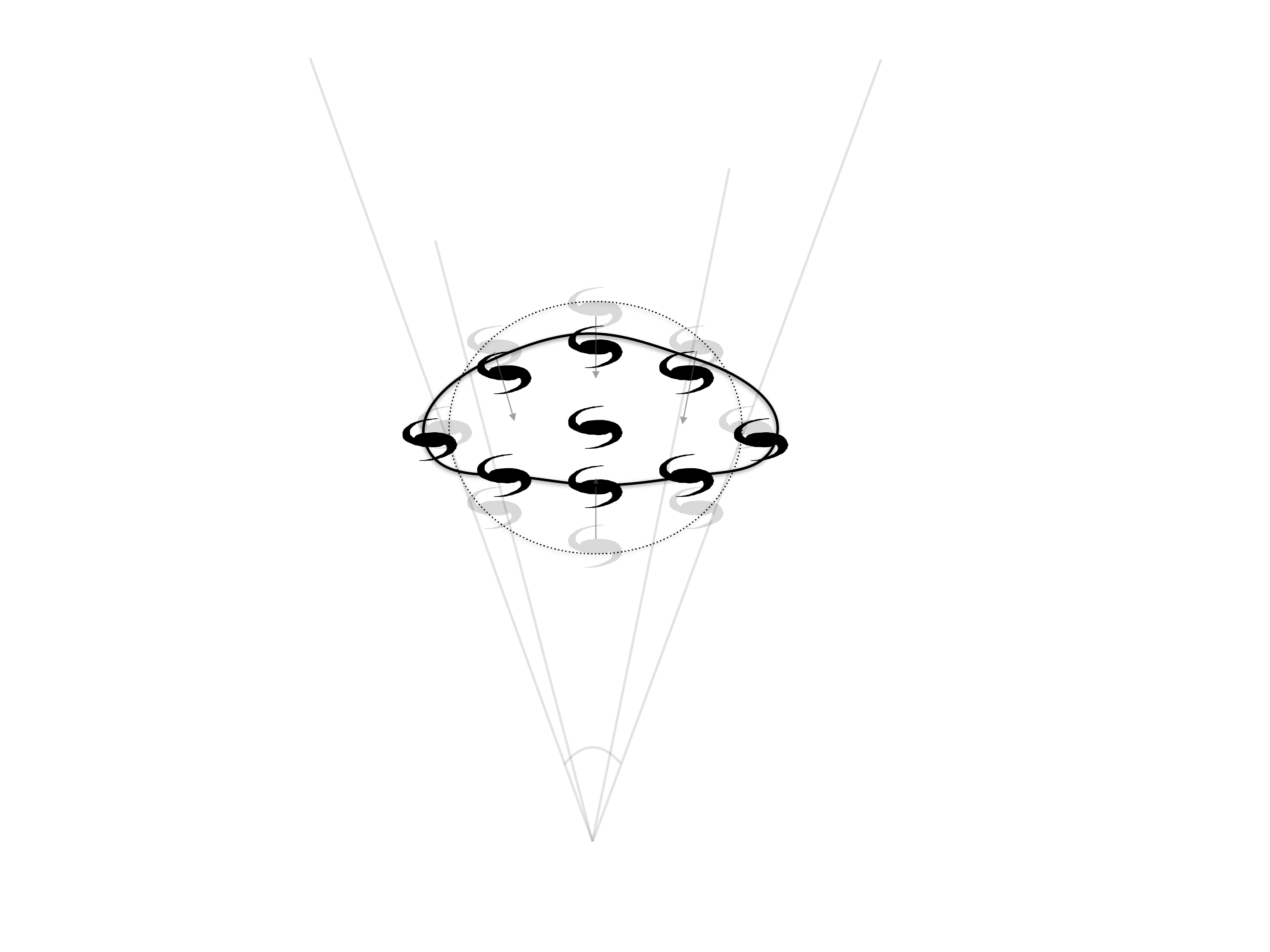}
\caption{{\it Upper panels}: Translation from real- to redshift- space of a galaxy pair in the transverse and non-transverse case, in the wide-angle case. Note that, even in the case of $|v_1| = |v_2|$, the presence of the observer changes the apparent scale {\bf s} in the transverse case, and both the apparent scale and the apparent angle w.r.t. the line of sight in the non-transverse case. {\it Bottom panels}: Large scale apparent modification of a spherical overdensity region. In the plane-parallel approximation (left panel), the Kaiser effect induces the so-called ``Pancakes of God'', so a spherical distribution of galaxies in real space will appear flattened in the radial direction in redshift space. In the wide-angle case (right panel), the introduction of an observer modifies the shape into a curved croissant-like shape that depends on the angular separation.
}
\label{fig:triangle}
\end{figure*}
\end{center}

We illustrate the physical reason for this in Figure~\ref{fig:triangle}.
The two plots in the top panel show that
the introduction of varying line of sights 
modifies the directions 
along which one moves galaxies when translating real-space to redshift-space positions. In the wide-angle picture, even galaxies with the same peculiar velocity will move along different line of sights, and the apparent pair separation and orientation angle will be different from the ones in real-space (and in the flat-sky approximation); this causes the so-called mode-coupling effects~\cite{Zaroubi:1993, Hamilton:1997, Raccanelli:2010}.
The two plots in the bottom panel illustrate the impact of this on the 
observed anisotropies in the galaxy clustering.
If we consider a spherical distribution of galaxies in real space, in the flat-sky approximation the large-scale Kaiser effect will cause the spherical distribution to appear flattened in the radial direction, giving origin to the so-called Pancakes of God (and on small scales to the Fingers of God). However, the introduction of several different line of sights (wide-angle description) and the Doppler term deform this shape, so that we can say the spherical distribution will be modified into {\it Croissants of God}. 
A pictorial representation is shown in the bottom panels of Figure~\ref{fig:triangle}, and more details about the physical effects described by the Doppler 
term will be discussed in Section~\ref{sec:alpha}.

\begin{center}
\begin{figure*}[htb!]
\includegraphics[width=0.49\columnwidth]{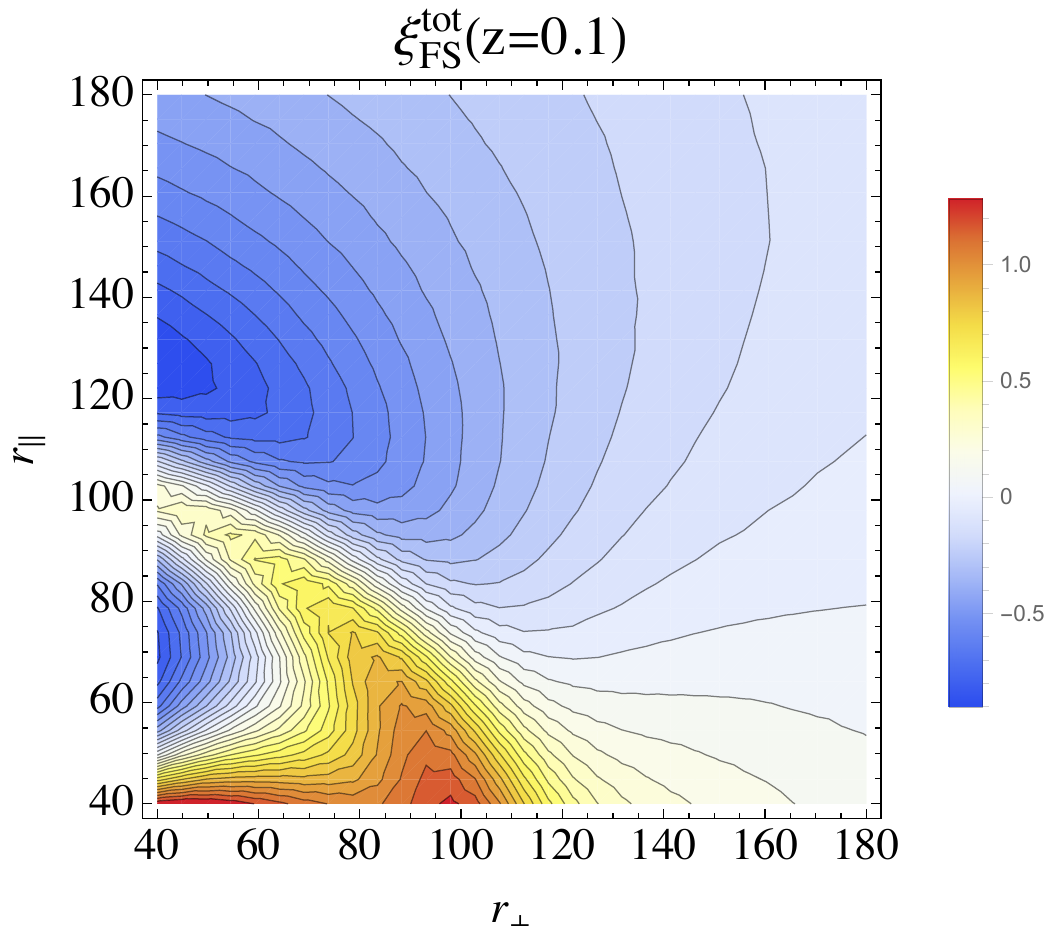}
\includegraphics[width=0.49\columnwidth]{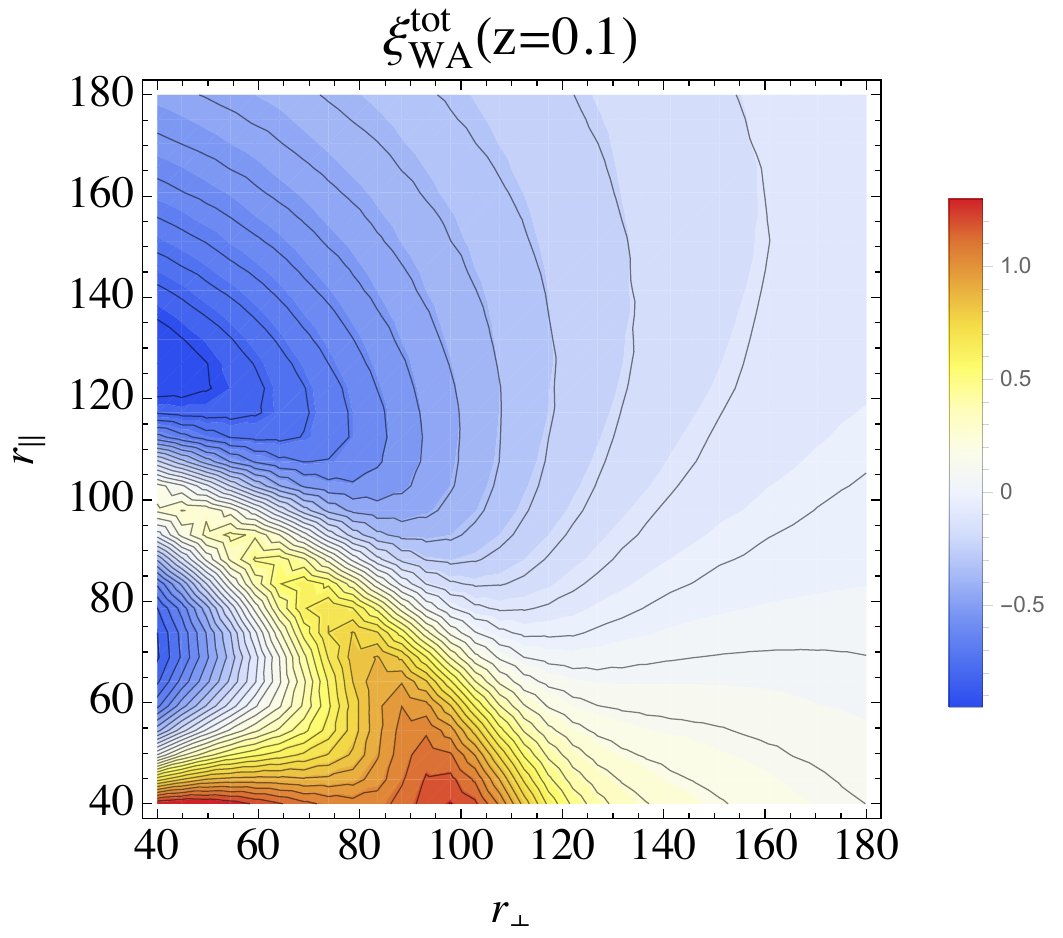}
\includegraphics[width=0.49\columnwidth]{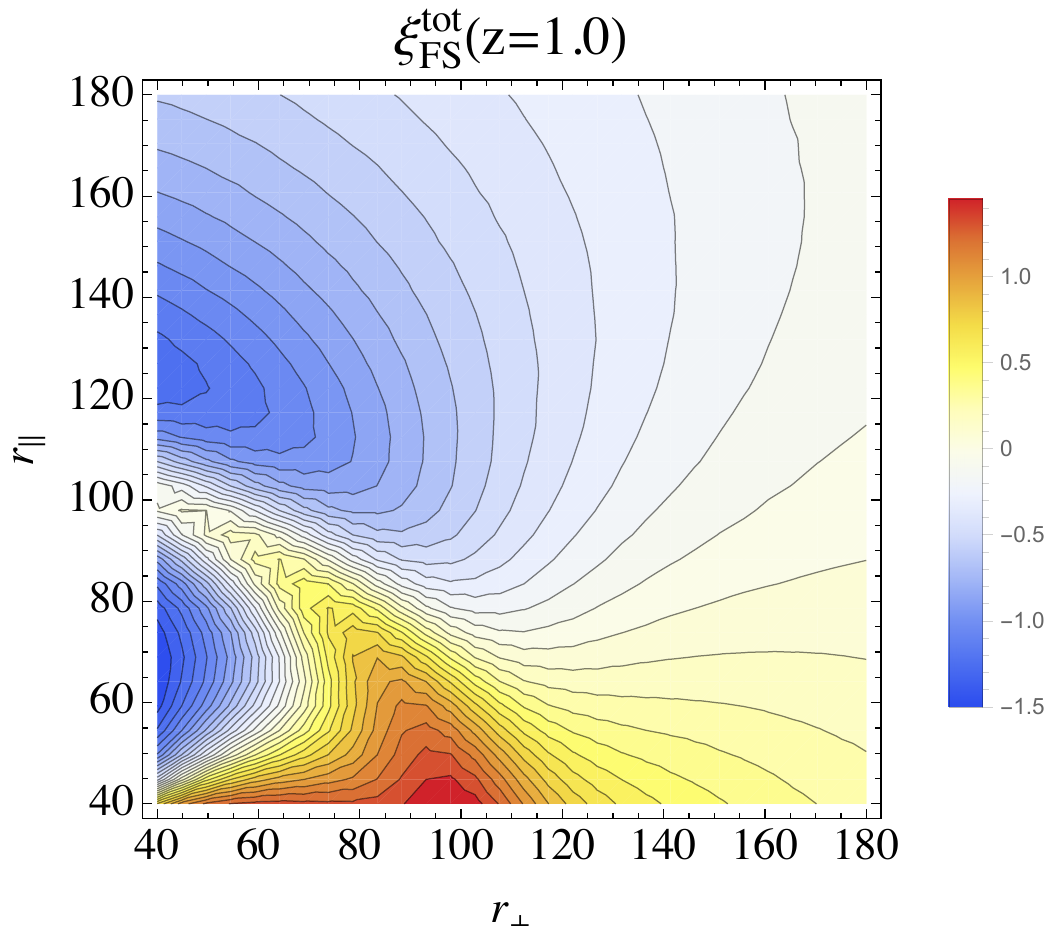}
\includegraphics[width=0.49\columnwidth]{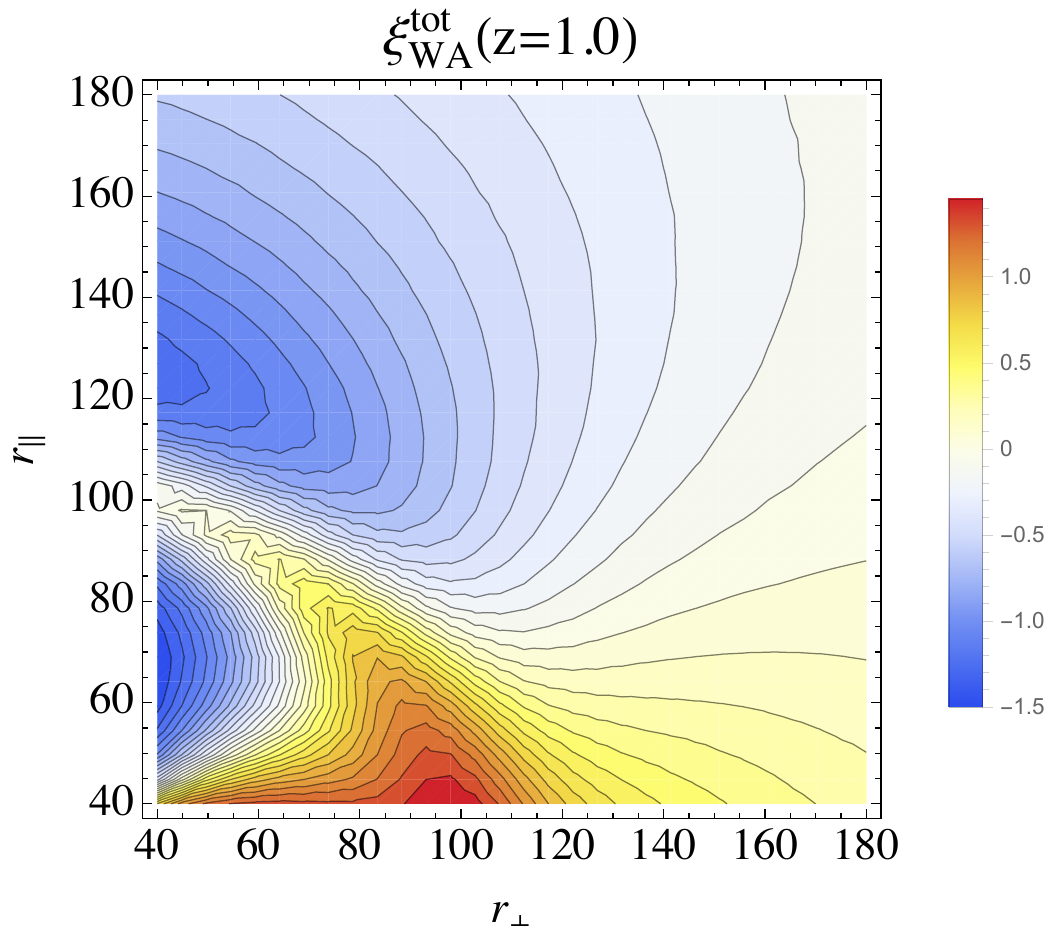}
\caption{{\it Top Left panel}: Galaxy correlation function at $z=0.1$ including
the Doppler term
(colored contours) compared with the $\xi^{\delta \delta}$ terms (black lines) in the flat-sky approximation. {\it Top Right panel}: Same as the top left panel, but when using the wide-angle formalism; {\it Bottom panels}: Same as above, but for $z=1$.
Here, and in subsequent plots, we plot ${\rm sinh}^{-1}(300\,\xi)$.
}
\label{fig:2D}
\end{figure*}
\end{center}

In Figure~\ref{fig:2D} we plot the 2D galaxy correlation function in linear theory, 
comparing the total correlation function with the correlation function 
without the Doppler term
in the flat-sky approximation and wide-angle formalisms. 
Colored contours indicate the full correlation, black lines the correlation excluding the Doppler term.
We can see how the effect of 
the Doppler term is suppressed in the flat-sky approximation and at 
high-z.

Black lines overlapping with borders between different colored contours indicates that the effect of including the Doppler term is negligible. At high redshift, this happens in both the plane-parallel and wide-angle cases (bottom panels). At low-z, in the plane-parallel approximation, the top left panel shows that Doppler terms are again negligible. However, the top right panel shows how the wide-angle low-z case clearly shows a discernible difference between the correlation function including the Doppler term and the model where it is neglected, and thus we can posit that the effect of
the Doppler term is not negligible. 
Throughout the paper, when plotting the 2D $\xi(r_\perp, r_\parallel)$, we will show ${\rm sinh}^{-1}(300 \, \xi)$, as in this way the correlation function is linear for small scales and logarithmic at large scales.
We will investigate in more detail the contribution of different terms in the following Section.

\section{Wide-angle galaxy correlation function}
\label{sec:xi_wa}
The galaxy two-point correlation function can be expressed using different 
formalisms. The fully wide-angle correlation function has been expressed 
using e.g. spherical harmonics expansion~\cite{Heavens:1995, Yoo:2013}, tripolar 
spherical harmonics~\cite{Szalay:1997, Papai:2008}, 
total angular momentum waves (TAM)~\cite{Dai:2012} and other formalisms~\cite{Tegmark:1995, Hamilton:1995, Zaroubi:1993}.

Here we use a particular case of the formalism of~\cite{Szalay:1997}, adapted to highlight and isolate the effect of the Doppler terms. To do so,
we write the galaxy correlation function as sum of three different contributions of density and Doppler terms:
\begin{align}
\label{eq:xi_tot}
\xi^{gg}(s,\theta,\varphi) = \xi^{<\delta \delta>}(s,\theta,\varphi)+\xi^{<\delta \alpha>}(s,\theta,\varphi)+\xi^{<\alpha \alpha>}(s,\theta,\varphi) \, .
\end{align}

We expand $\xi$ using a combination of trigonometric functions in the case where $s$ is the linear separation (in redshift-space) between the galaxies, $\theta$ is the angular separation (i.e. the angular aperture in Figure 1) and $\varphi$ the pair orientation angle (i.e. the angle between the $s$ linear separation and the bisector of $\theta$). We fix $\varphi = 0$ for the radial direction (galaxies behind each other); see also~\cite{Szalay:1997, Papai:2008} for alternative choices of coordinates and expansions, and a recent detailed discussion can be found in~\cite{Slepian:2015}.

We find the following expressions for the $<\delta \delta>$, $<\delta \alpha>$, $<\alpha \alpha>$ components:
\begin{align}
\label{eq:xi_ii}
\xi^{\delta \delta}(s,\theta,\varphi) = 
\xi_0^2&\left(\frac{1}{15}\right)\left[15+10 f +  f^2 +2 f^2 \cos^2 2\theta \right]+ 
\\
-\xi_2^2 &\left( \frac{f}{42}\right) \left[14+2f+4f \cos^2 2\theta + (21+9f) \left(\cos 2\theta+\cos 2\varphi \right) \right]
\nonumber\\
+\xi_4^2& \left(\frac{f^2}{280}\right) \left[6+3\cos 4\theta+35\cos 4\varphi +20(\cos 2\theta+\cos 2\varphi)\right]
 \, ; \nonumber \\ \nonumber
\xi^{\delta \alpha}
=\frac{2 \alpha}{s(\cos 2 \varphi - \cos 2 \theta)}&  
\Bigg\{ \xi_1^1\  f  \sin^2 2\theta  \left[ (1+  \frac{f}{5})\cos 2\varphi +  \frac{2f}{5} \cos 2\theta \right] 
\\ 
+ \xi_3^1\ &\Bigg[-\frac{ f^2}{40}\left[ -3\cos 2\theta +4\cos^3 2\theta+ 12 \cos 2\varphi \sin^2  2\theta \right] \Bigg]\Bigg\}\nonumber
 \, ; \nonumber \\
\xi^{\alpha \alpha}(s,\theta,\varphi) 
&= \frac{\alpha^2}{s^2}  \frac{\sin^2 2 \theta}{\cos 2 \varphi - \cos 2 \theta}
 \Bigg\{ \xi_0^0 \left[ \frac{4}{3} f^2 \cos 2\theta  \right] + 
\xi_2^0 \left[ -\frac{2}{3} f^2 \left(\cos 2\theta +3\cos 2\varphi \right) \right] \Bigg\} \, .
\end{align}
Here:
\begin{equation}
\xi_{\ell}^m(s) 
= \int \frac{dk}{2\pi^2} k^{2-m} j_\ell(s k) \, P_\delta(k) \, ,
\end{equation} 
are spherical Bessel transforms of the matter power spectrum~\cite{Papai:2008, Raccanelli:2010, Raccanelli:3D} and $\alpha$, as we will see in the next Section, is a function of redshift and other cosmological parameters.
We verified that our Equations~\eqref{eq:xi_ii} are perfectly consistent with the results of~\cite{Szalay:1997, Papai:2008, Raccanelli:2010}.
In Appendix~\ref{app:xi} we illustrate their derivation.

\begin{center}
\begin{figure*}[htb!]
\includegraphics[width=0.49\columnwidth]{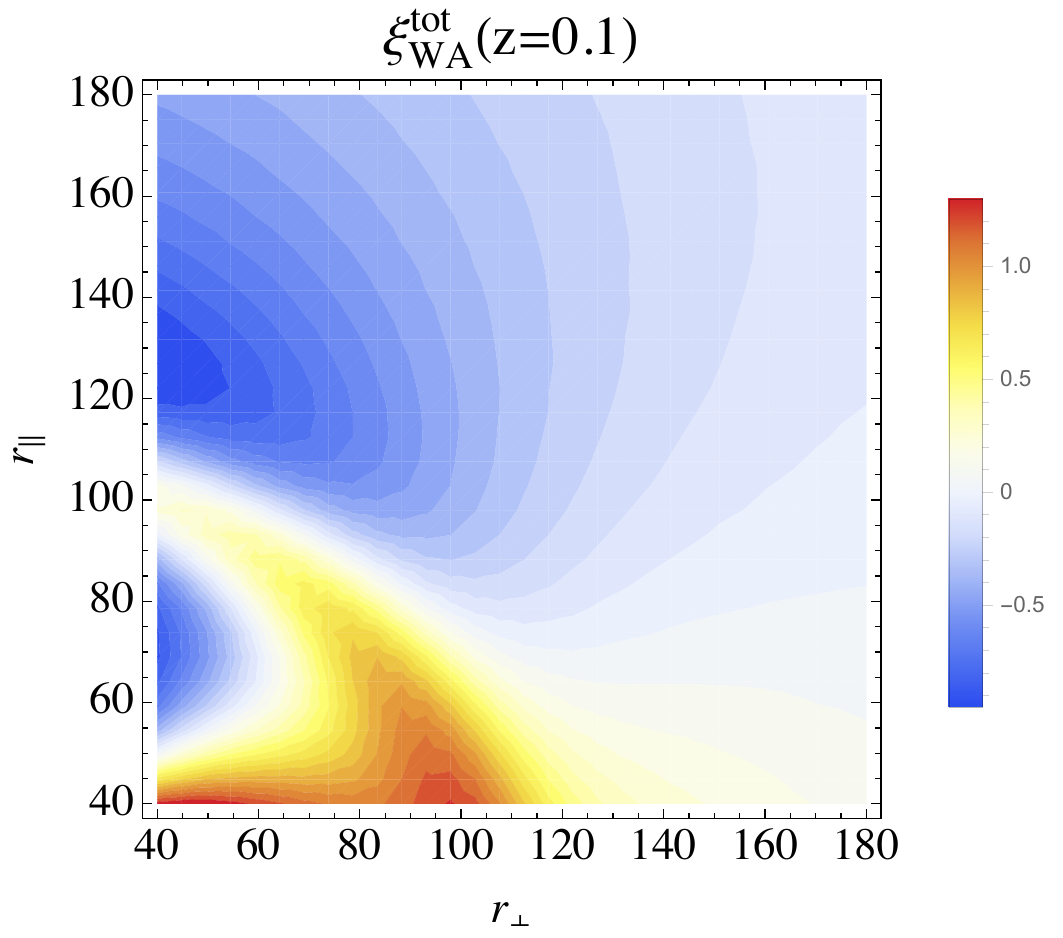}
\includegraphics[width=0.49\columnwidth]{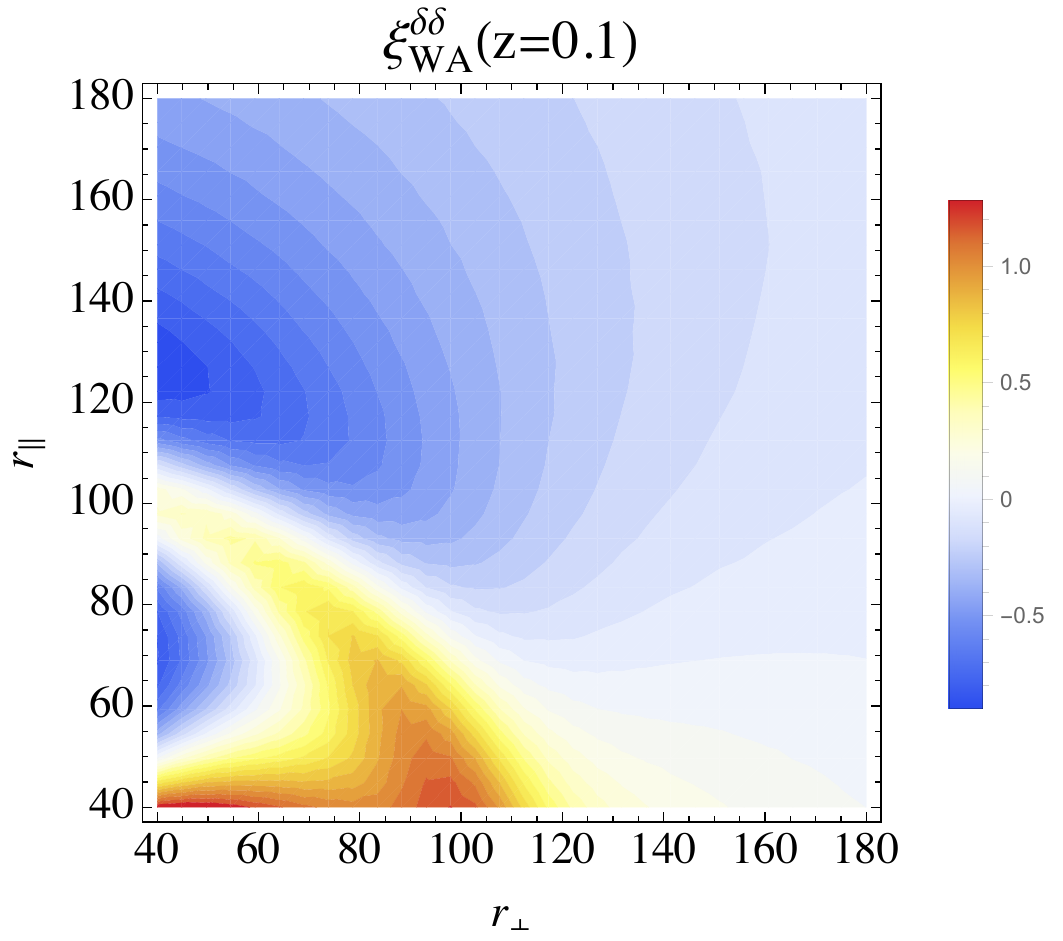}
\includegraphics[width=0.49\columnwidth]{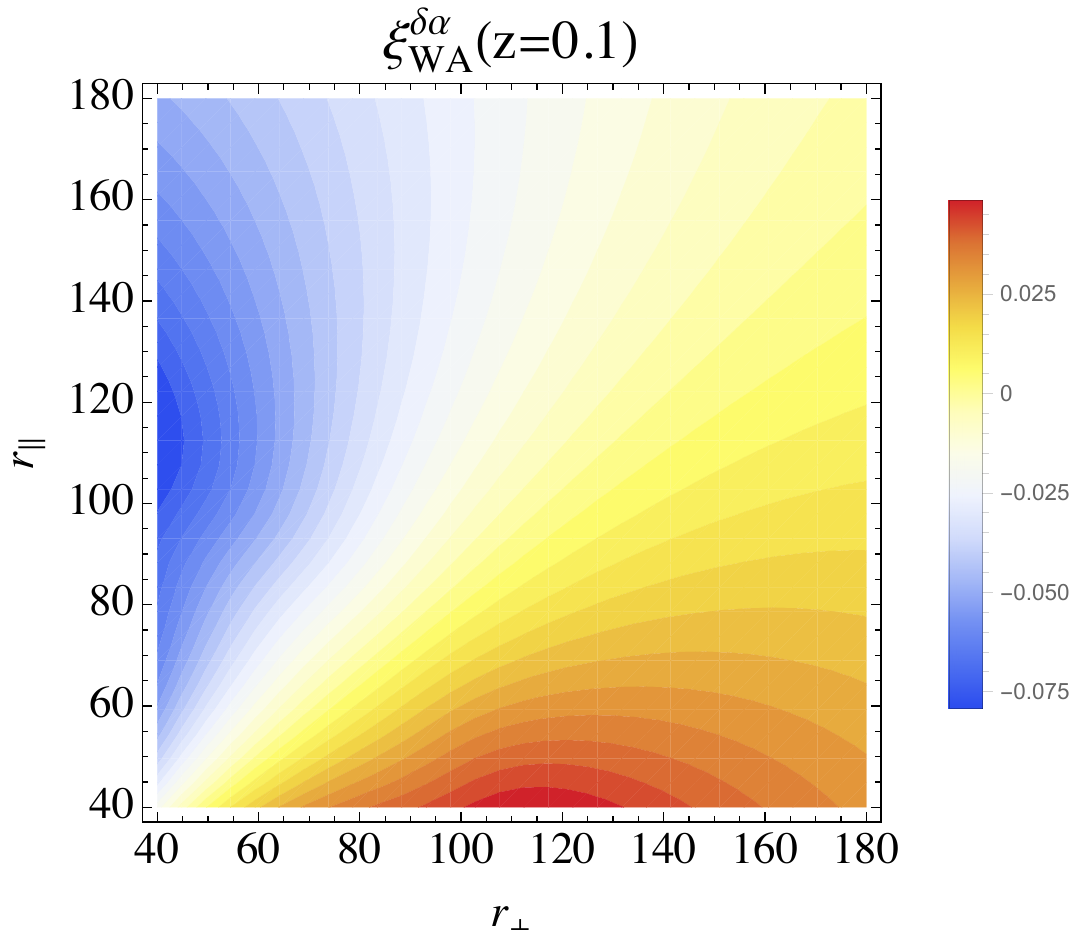}
\includegraphics[width=0.49\columnwidth]{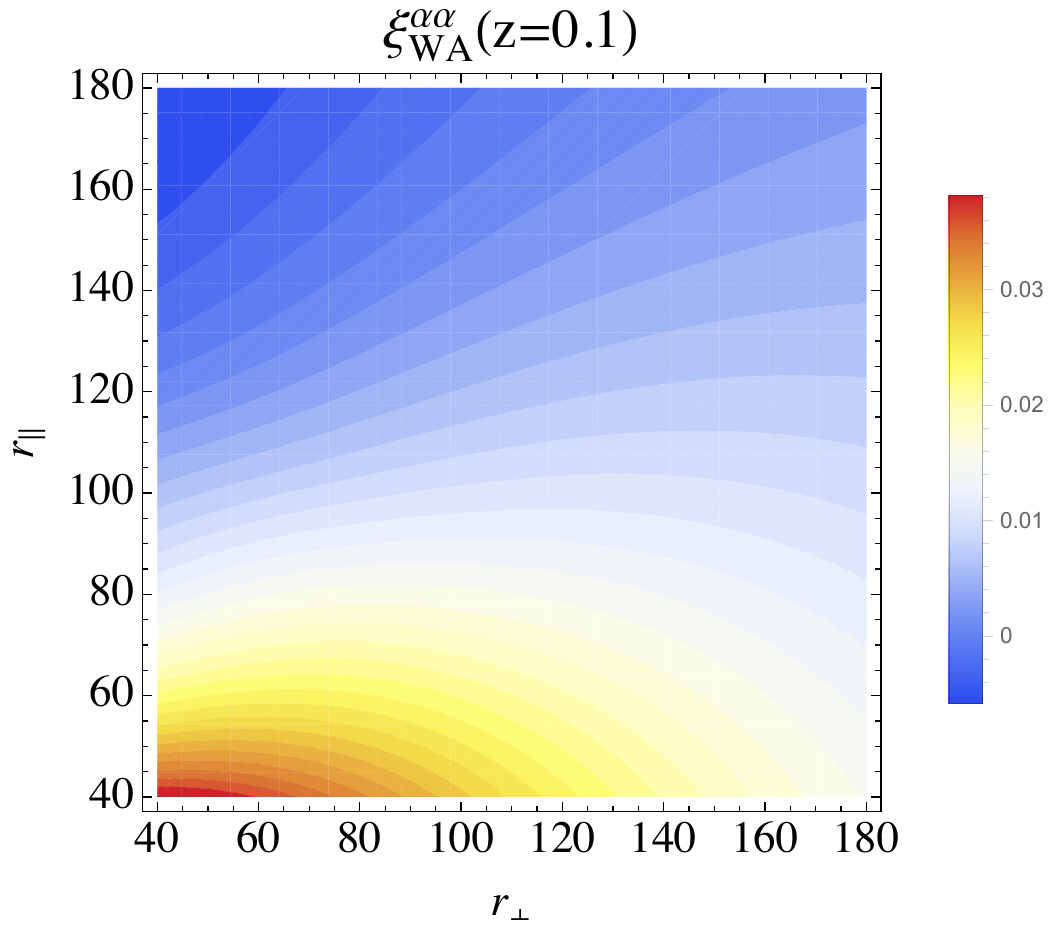}
\caption{2D redshift-space galaxy correlation function including wide-angle terms, for the models including all terms (top left panel),  $\langle \delta \delta \rangle$ (top right panel), $\langle \delta \alpha \rangle$ (bottom left panel),  $\langle \alpha \alpha \rangle$ (bottom right panel). Results here are shown for $\alpha=2$, to focus on the wide-angle geometrical contribution.}
\label{fig:2D_comp}
\end{figure*}
\end{center}

In Figure~\ref{fig:2D_comp} we show the different contributions to the 2D galaxy correlation function, in linear theory. We plot, using the wide-angle formalism of Equations~\ref{eq:xi_ii}, the total, 
$\xi^{\delta\delta}$, $\xi^{\alpha\alpha}$, and $\xi^{\delta\alpha}$ terms.
Results are shown at $z=0.1$; we use here a constant $\alpha=2$ for simplicity, but its value will be redshift- and (strongly) survey dependent. This is to show the relevance of the geometrical effects. Assuming $\alpha=2$ is equivalent to assuming a volume limited survey sampling the population of galaxy whose comoving density stays the same, and Doppler lensing effects are suppressed; their importance will be investigated in more detail in Section~\ref{sec:snr}.
It can be seen that the magnitude of both the pure mode-coupling ($\langle \alpha \alpha \rangle$) and the mixed ($\langle \delta \alpha \rangle$) terms is larger for transverse and radial correlations, and that the sum of the two can be of order 5\% of the full correlation. Considering the precision in future measurements of galaxy clustering, this correction could be relevant.

A particular case that is worth mentioning is when the two lines of sight are orthogonal, $2 \theta = \pi/4$, where Equations~\ref{eq:xi_ii} are greatly simplified.
Here, $\cos 2 \theta = 0, \sin 2 \theta=1$.
One can see that the $\varphi$ dependence only shows up in the $\delta\delta$ terms, not in the other two, as expected.
In this case, we can rewrite the correlation function as:
\begin{align}
\xi^{\delta \delta} =&\frac{\xi_0^2}{15}\left[15+10 f +  f^2  \right]
-\frac{\xi_2^2}{42} f \left[14+2f + (21+9f) \cos 2\varphi \right]
+\frac{f^2\xi_4^2 }{280} \left[3+35\cos 4\varphi +20 \cos 2\varphi \right]
\\
\xi^{\delta \alpha}
=&\frac{2 \alpha f}{s} \Bigg[
\xi_1^1\   (1+  \frac{f}{5}) -\xi_3^1\ \left(\frac{3 f}{10} \right)
\Bigg]
\\
\xi^{\alpha \alpha}
=& -\frac{2 \alpha^2 f^2}{s^2}  
 \xi_2^0 \, .
\end{align}
In this configuration, the separation angle is very large, so we expect mode-coupling effects to be also very strong; we also notice that the cross term is now independent on the pair orientation angle $\varphi$.
While it may seem that the limit $\theta \rightarrow 0$ is singular due to the $g$ angular terms, we checked that $\xi$ is well behaved. For illustrative purposes, here we write the expansion around $\theta=0$ for the three components:
\begin{align}
\label{eq:xi_ii_0}
\xi^{\delta \delta}(s,\theta\rightarrow0,\varphi) &= \left[ 1+\frac{2}{3}f+\frac{1}{5}f^2 \right] \xi_0^2 +
\frac{f}{42}\left[-14 -42 \cos(2\varphi) -6f -18 \cos(2\varphi)f \right] \xi_2^2 + \nonumber \\
& + \frac{1}{280}\left[9+20\cos(2\varphi)+35\cos(4\varphi) f^2 \right] \xi_4^2 \, ; \\ \nonumber
\xi^{\delta \alpha}(s,\theta\rightarrow0,\varphi) &=
-\frac{2\alpha}{5s} \Bigg\{ \left[10 \cos(2\varphi) \csc(\varphi)^2 f + 4 \csc(\varphi)^2 f^2 +2\cos(2\varphi) \csc(\varphi)^2 f^2 \right]\xi_1^1 - \nonumber \\
& - \left[\csc(\varphi)^2 f^2 - 3\cos(2\varphi) \csc(\varphi)^2 f^2\right]\xi_3^1 \Bigg\} \, ; \nonumber \\
\xi^{\alpha \alpha}(s,\theta\rightarrow0,\varphi) &= \frac{2}{3s^2} \Bigg\{ \csc(\varphi)^2\alpha^2
\left[2f^2 \xi_0^0 - (f^2-3\cos(2\varphi)f^2) \xi_2^0 \right] \Bigg\} \, .
\end{align}

\section{Velocity and Doppler terms: $\alpha$}
\label{sec:alpha}

As we saw in Section~\ref{sec:fs_wa}, the full expression 
relating real- to redshift- space includes terms involving $\alpha$, as defined in Eq.\ (\ref{eq:alphaN}); these originate from the part of the Jacobian proportional to $v_r/r$. We call 
these {\it Doppler terms}.
They arise from the combination of two effects: 
a geometrical modification to the RSD operator and the Doppler lensing.

It has been shown in the past that these terms induce a ``mode leaking" 
(they are sometimes also called ``mode-coupling" in 
literature~\cite{Zaroubi:1993}), because the full RSD operator ${\bf S}$ of Equation~\eqref{eq:kaiser_operator} destroys 
the three-dimensional translational symmetry, so that Fourier modes are 
no longer eigenmodes of the distortion operator, and the redshift space
two point correlator in Fourier space is no longer proportional to $\delta^{\rm D}(k-k')$.
Therefore, clustering information ``leaks" from the real-space scale to a (modified) redshift-space one (this can be seen from the top panels of Figure~\ref{fig:triangle}).
The redshift distortion operator in this case is not Hermitian, unlike the plane-parallel one. This means that the eigenfunctions of the RSD operator do not all have real eigenvalues, and so ${\bf S}$ causes modes close to the observer to undergo a phase shift in passing from real to redshift space.
These ``geometry" terms account for the fact that galaxy pairs coherently move to the high-density from low-density regions~\cite{Raccanelli:2010}; this can be seen also from the fact \aaa depends on $\ud \, N(z) / \, \ud z$.

The Doppler term 
has been derived in the Newtonian~\cite{Hamilton:1997, Szalay:1997, Szapudi:2004, Papai:2008, Raccanelli:2010} and general relativistic case~\cite{Jeong:2012, Bertacca:2012}; investigations on the contribution of the 
Doppler term can be also found in e.g.~\cite{DiDio:2013CLASS, Bonvin:2014}.

The Newtonian form of $\alpha$ of Equation~\eqref{eq:alphaN} can be found in e.g.~\cite{Hamilton:1997}; here we follow the (equivalent) definition given in~\cite{Szalay:1997}.

\begin{center}
\begin{figure*}[htb!]
\includegraphics[width=0.49\columnwidth]{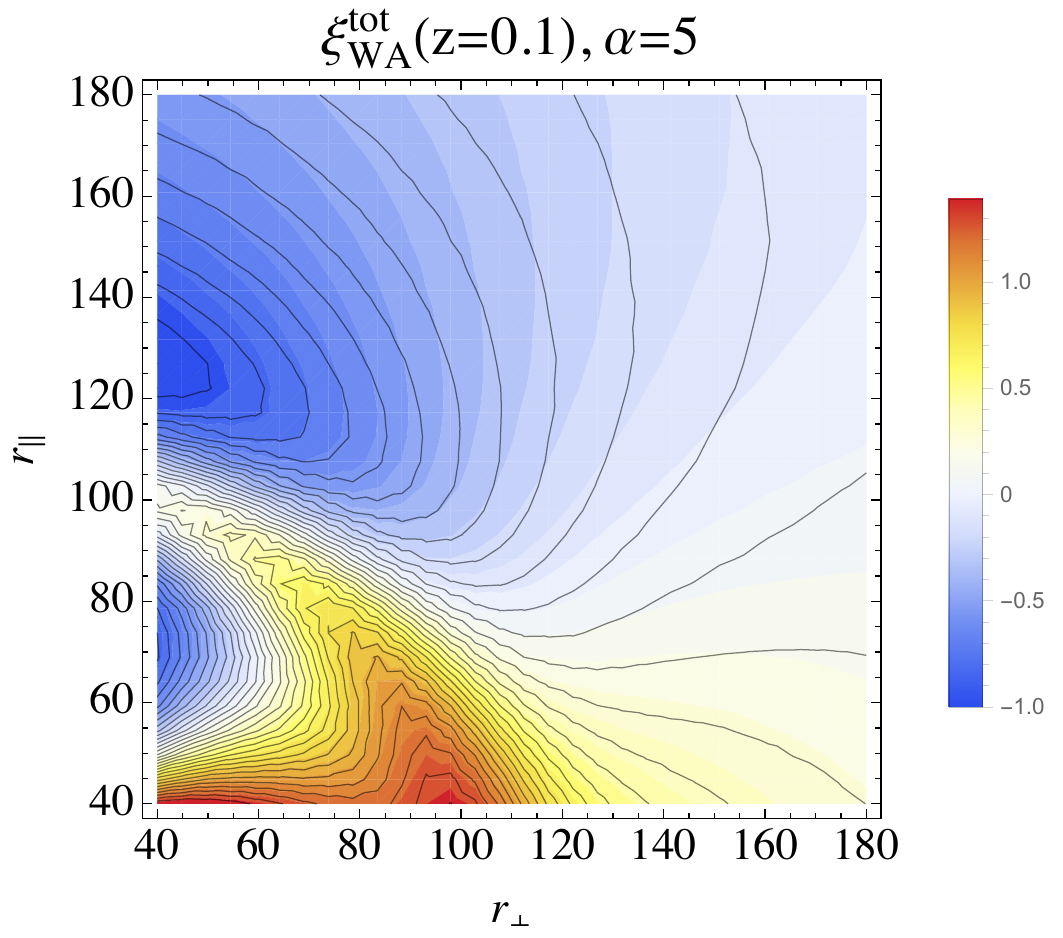}
\includegraphics[width=0.49\columnwidth]{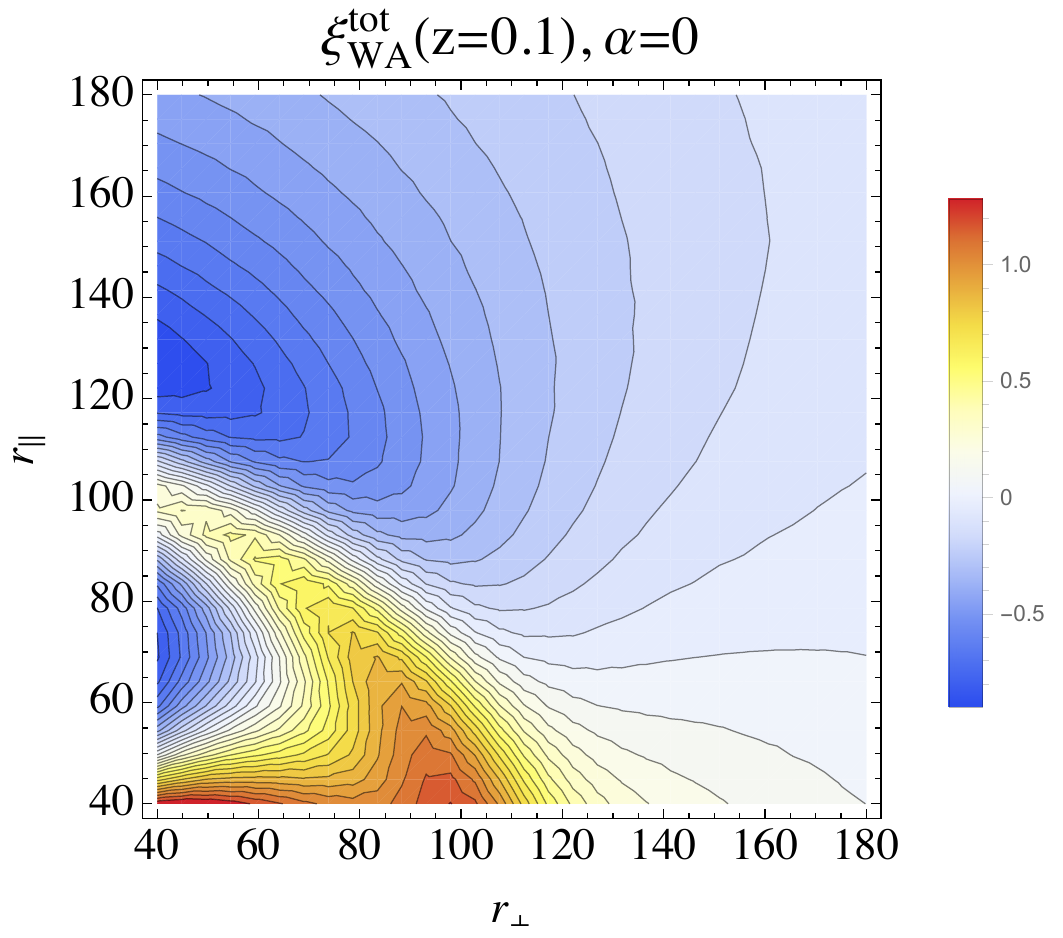}
\caption{Same as the top right panel of Figure~\ref{fig:2D} ($z=0.1$, wide-angle), but for $\alpha=5$ ({\it left panel}) and $\alpha=0$ ({\it right panel}).
}
\label{fig:2D_alpha}
\end{figure*}
\end{center}

In Figure~\ref{fig:2D_alpha} we show the effect (in the same way shown in Figure~\ref{fig:2D}) of $\alpha=0$ and $\alpha=5$ which roughly corresponds to the value obtained from a gaussian galaxy distribution centered at $z=0.1$ and with $\sigma=0.1$. As expected, the deviation from the $\langle \delta \delta \rangle$ case increases with $\alpha$.

The Doppler lensing terms account for apparent modifications of angular size 
and magnitude of a galaxy, due to the motion of the source galaxy with respect to 
the observer. This effect originates from the aberration effect in special relativity.
Galaxies with the same (apparent) redshift space position can be, in real space, more distant (if they are moving toward us, $V \cdot \hat{n} > 0$), than galaxies with no peculiar velocity, or closer to us (if $V \cdot \hat{n} < 0$).

This generates two opposite effects on the observed angular size of the galaxy: a galaxy that is more distant in real space is observed under a smaller solid angle, but its photons were emitted at an earlier time, when the value of the scale factor of the universe was smaller.
Hence, those photons experienced a larger stretch in their path toward us, which increases the observed angular size of the galaxy.
A study of the Doppler lensing effect in the context of lensing analyses can be found in~\cite{Bacon:2014}, where it was also shown that at low z Doppler lensing effects dominate even over cosmic magnification, as they are enhanced by a factor $1/(aHr)$.

In the general form, including relativistic corrections, $\alpha$ can be written as~\cite{Bertacca:2012}:
\begin{eqnarray}
\label{eq:alpha}
\alpha(z)&=& - \chi(z) \frac{H(z)}{(1+z)} \left[b_e(z) -
1-2\mathcal{Q}(z) +\frac{3}{2}\Omega_m (z)-
\frac{2}{\chi(z)}\big[1-\mathcal{Q}(z)\big]\frac{(1+z)}{H(z)}\right] \; ,
\end{eqnarray}
where the magnification bias parameter ${\cal Q}$ for a magnitude-limited survey is \cite{Geach:2009, Jeong:2012}:
\begin{equation}
\label{eq:q}
\mathcal{Q} = {d \ln N_g \over d \ln \mathcal{L}}\bigg|_{{\cal L}={\cal L}_{\rm lim}}, 
\end{equation}
and $N_g$ is the comoving number density of galaxies of luminosity $\mathcal{L}>{\cal L}_{\rm lim}$~\cite{Liu:2014}. It is worth noting that in literature the magnification bias is sometimes indicated as $s(z)$, and $\mathcal{Q}=5s/2$.
It can be verified that the newtonian form of $\alpha$ can be obtained if we rewrite Equation~\eqref{eq:alpha} as:
\begin{eqnarray}\label{alpha2}
\frac{\alpha(z)}{\chi(z)} =
-  \frac{H(z)}{(1+z)} \left[\frac{3}{2}\Omega_m (z)-1\right] + \frac{d
\ln{n_g}}{d \chi}+ \frac{2}{\chi} \, ;
\end{eqnarray}
$\chi(z)$ is the comoving distance, $b(z)$ is the galaxy bias and:
\begin{equation}
b_e ( {z})=-(1+ {z}) {d\ln  [n_g(1+z)^{-3}] \over d z} \, .
\end{equation}

It is interesting to note that the $velocity$ terms depend on $\alpha$, and the relativistic expression for $\alpha$ depends on the time derivative of $aH$, 
hence allowing a new way of measuring the derivative of the Hubble parameter,
or {\it cosmic acceleration}.
We can isolate different contributions to \aaa by diving it into $\alpha = \alpha_1 + \alpha_2 + \alpha_3$, and rewriting Equation~\eqref{eq:alpha} as:
\begin{eqnarray}
\label{eq:alpha_i}
\alpha_1 = 2 - b_e \frac{H(z) \chi(z)}{(1+z)} \, ; \,\,\,\,\,
\alpha_2 = 2 \mathcal{Q}(z) \left[\frac{H(z)\chi(z)}{(1+z)}-1\right] \, ; \,\,\,\,\,
\alpha_3 = \frac{H(z) \chi(z)}{(1+z)} \left[1-\frac{3}{2}\Omega_m(z)\right] \; .
\end{eqnarray}
Here we separate the geometry and mode coupling terms ($\alpha_1$), the Doppler lensing ($\alpha_2$) and the acceleration terms ($\alpha_3$). This term effectively represents a volume distortion of the bin observed. Being able to measure $\alpha_3$ would then represent a direct measurement of cosmic acceleration. Unfortunately this term is subdominant, so a precise detection will be difficult to obtain. We leave to a future work an investigation on survey requirements needed to obtain such a measurement.

From a physical point of view, these effects are important at wide angular separations because the signal we are looking for is subdominant if compared to the gravitational growth. By looking at wide angular separations, we are considering two line of sights that are mostly gravitationally uncorrelated, and so these Doppler terms can become more important. Then one could wonder why it is convenient to use the 2-point correlation function as a probe to measure those effects: the answer is that we are measuring the variance of a clumped distribution, and so the second moment is the appropriate statistics to measure. A more detailed investigation of alternative and possibly more optimal ways to measure velocity and Doppler terms is left for a future work.

\section{Detection of velocity and Doppler terms}
\label{sec:snr}
In this Section we try to understand if mode-coupling, velocity and Doppler terms could be detected using future galaxy redshift surveys. 
We focus our analysis on measurements of the power spectrum and angular spectra.
We will perform an analysis of how to measure the Doppler term in configuration space, and potentially use them for cosmological measurements, using N-body simulations, in a companion paper in preparation.

\subsection{Plane-parallel power spectrum}
We start by writing the plane-parallel power spectrum including the terms 
proportional to \aaa in Equation~\eqref{eq:J}. Following the approach of~\cite{Papai:2008, Raccanelli:2010}, and assuming the flat-sky approximation, we can rewrite the RSD operator of Equation~\eqref{eq:kaiser_operator} by adding the term (see~\cite{Bertacca:2012} for a similar derivation in configuration space, including relativistic effects):
\begin{equation}
{\mathbf S}^{\rm tot} = {\mathbf S}^{\rm Kaiser}+\text{\textphnc{\Ashin}}_\alpha \, , \,\,\, \text{\textphnc{\Ashin}}_\alpha = \alpha \frac{\beta \mu}{k\chi} \, .
\end{equation}
so that the redshift-space power spectrum will be:
\begin{equation}
\label{eq:pk_a}
P^s(k,\mu) = \left[ \left(1 + \beta \mu^2 \right)^2 + \left( \alpha \frac{\beta \mu}{k\chi}\right)^2 \right] \, P^r(k) \, .
\end{equation}
Here $\chi(z)$ is the distance between the observer and the survey volume, and the other quantities have the usual meaning.

We can then generalize the standard practice of expanding the power spectrum as a sum of Legendre polynomials, and we obtain:
\begin{eqnarray}
\label{eq:Pk_alpha}
P^s(k,\mu) &=& \sum_{\ell=0}^{\ell=4} P^r(k) A_\ell (k) \mathcal{L}_\ell(\mu) \, ,
\end{eqnarray}
where $\mathcal{L}_\ell$ are Legendre polynomials of the cosine of the pair orientation angle $\varphi$ ($\mu =$ Cos$(\varphi)$), and:
\begin{eqnarray}
A_0 &=& 1+ \frac{2}{3}\beta + \frac{1}{5}\beta^2 + \frac{\alpha^2}{3k^2 \chi^2} \beta^2 \, ; \\
A_2 &=& \frac{4}{3}\beta + \frac{4}{7}\beta^2 + \frac{2\alpha^2}{3k^2 \chi^2} \beta^2 \, ; \nonumber \\
A_4 &=& \frac{8}{35}\beta^2 \; . \nonumber
\end{eqnarray}
Once again we can see how the term describing the effect of Doppler contains an additional $k$ dependence, causing the phase shifts when going from the real- to the redshift- space.
Odd multipoles vanish in this approximation (see~\cite{Hamilton:1992}), but they are present in the full 3D description of galaxy clustering~\cite{Raccanelli:3D, Bonvin:2014}.

\begin{center}
\begin{figure*}[htb!]
\includegraphics[width=0.49\columnwidth]{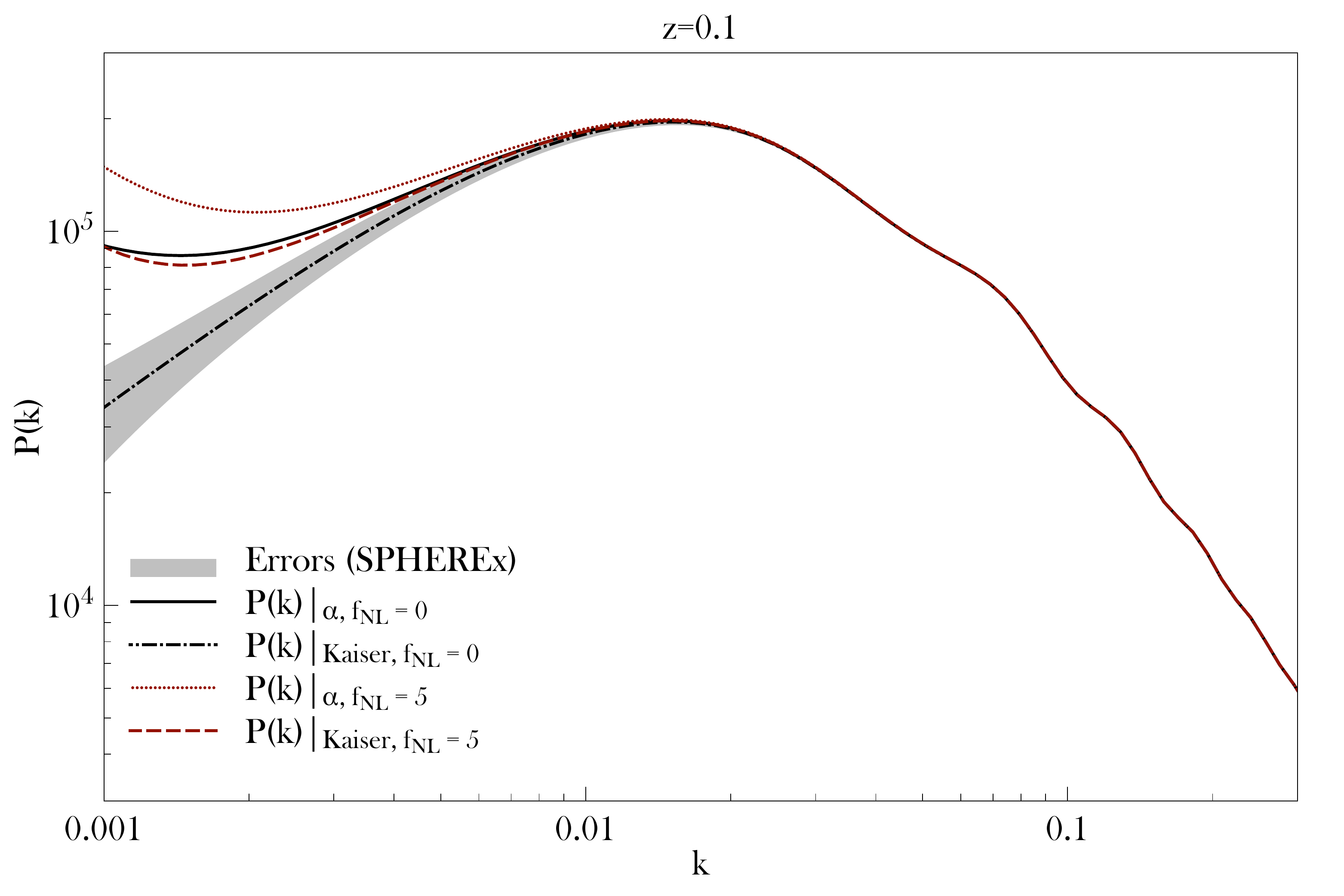}
\includegraphics[width=0.49\columnwidth]{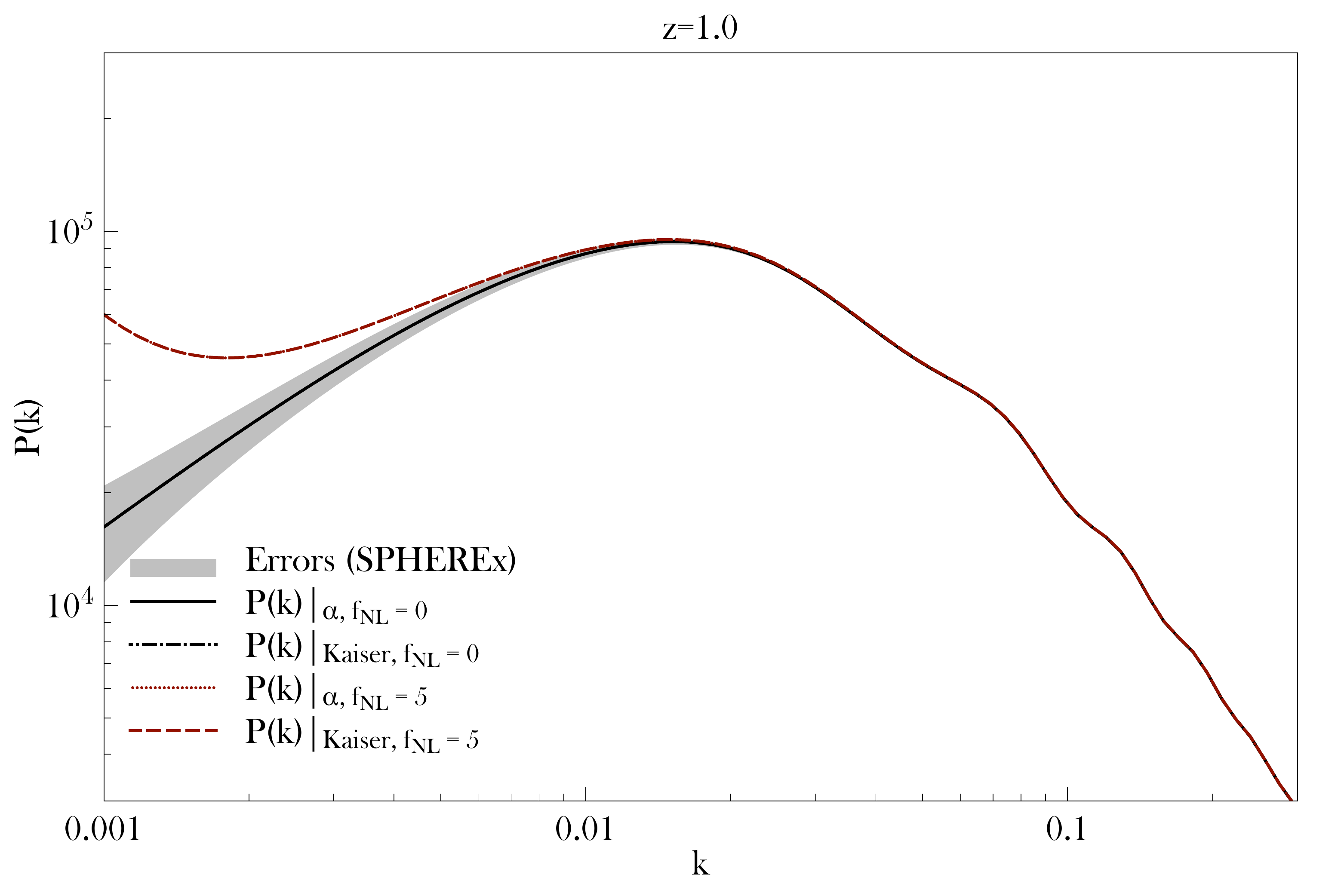}
\caption{Angle averaged power spectrum, at $z=0.1$ (left panel), and $z=1.0$ (right panel).
Lines show the power spectrum for the Kaiser formalism with no non-Gaussianity (black dot-dashed) and when including Doppler terms (black solid).
Red lines show the case $f_{\rm NL}=5$, in the Kaiser (dashed) and Kaiser+Doppler (dotted) cases. Including Doppler terms mimics an effective $f_{\rm NL}^{\rm loc}$ at low-z.}
\label{fig:Pk_alpha}
\end{figure*}
\end{center}

In Figure~\ref{fig:Pk_alpha} we show the effect of adding the operator $\text{\textphnc{\Ashin}}_\alpha$.
We can see that in the low-z regime, even in the plane-parallel approximation, Doppler terms mimics the effect of a local non-Gaussianity with $f_{\rm NL}^{\rm loc}$ of a few. This result is somewhat similar to the one of~\cite{Jeong:2012}; however, here we do not include gravitational potential and lensing terms. This means that, at low redshift, the Doppler terms represent the major contribution to (local) large-scale effects, if the magnification bias is large enough (more details on the magnification bias and its impact on the magnitude of Doppler terms are given in Section~\ref{sec:cls}.

\subsubsection{Effects on parameter estimation}
\begin{center}
\begin{figure*}[htb!]
\includegraphics[width=0.49\columnwidth]{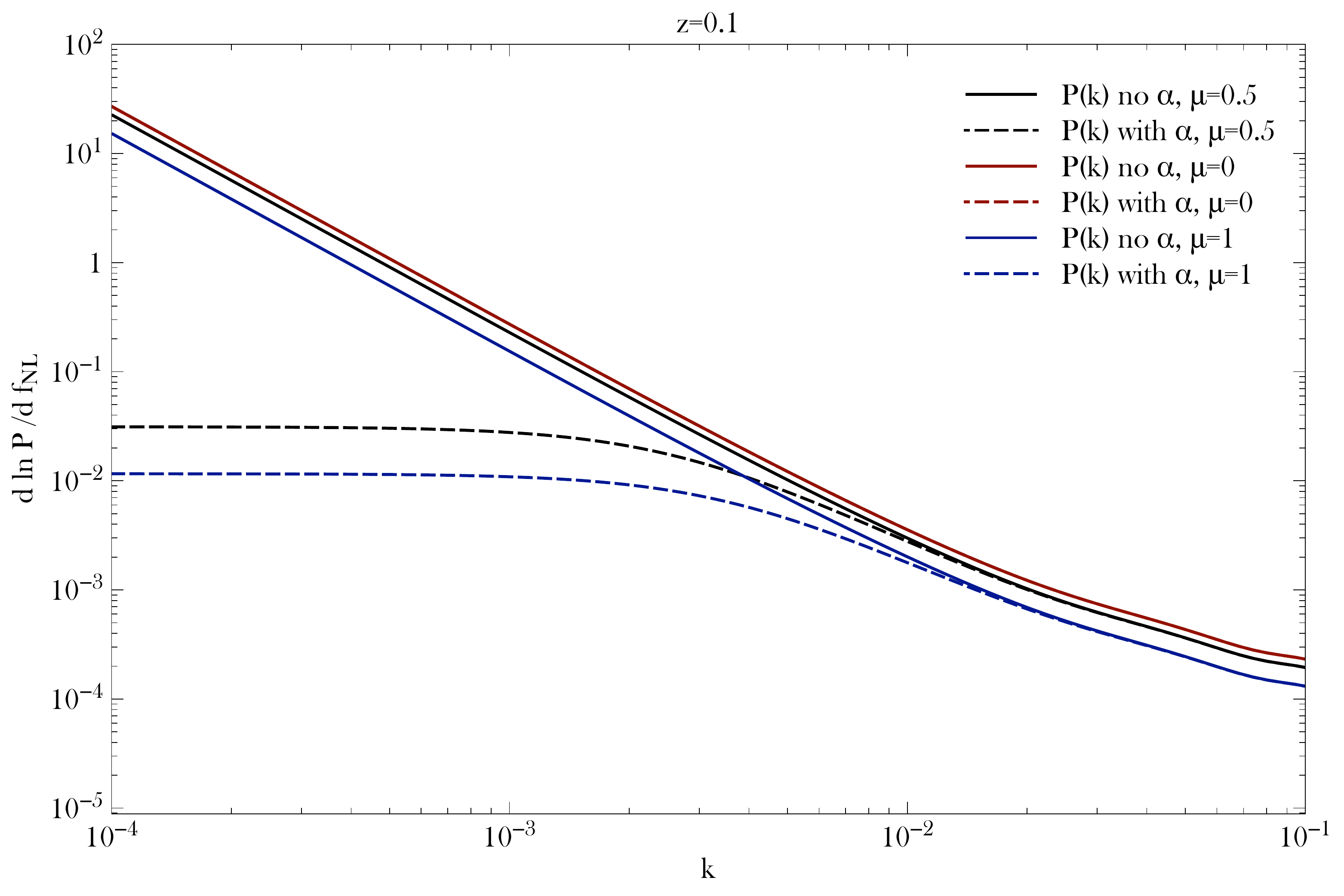}
\includegraphics[width=0.49\columnwidth]{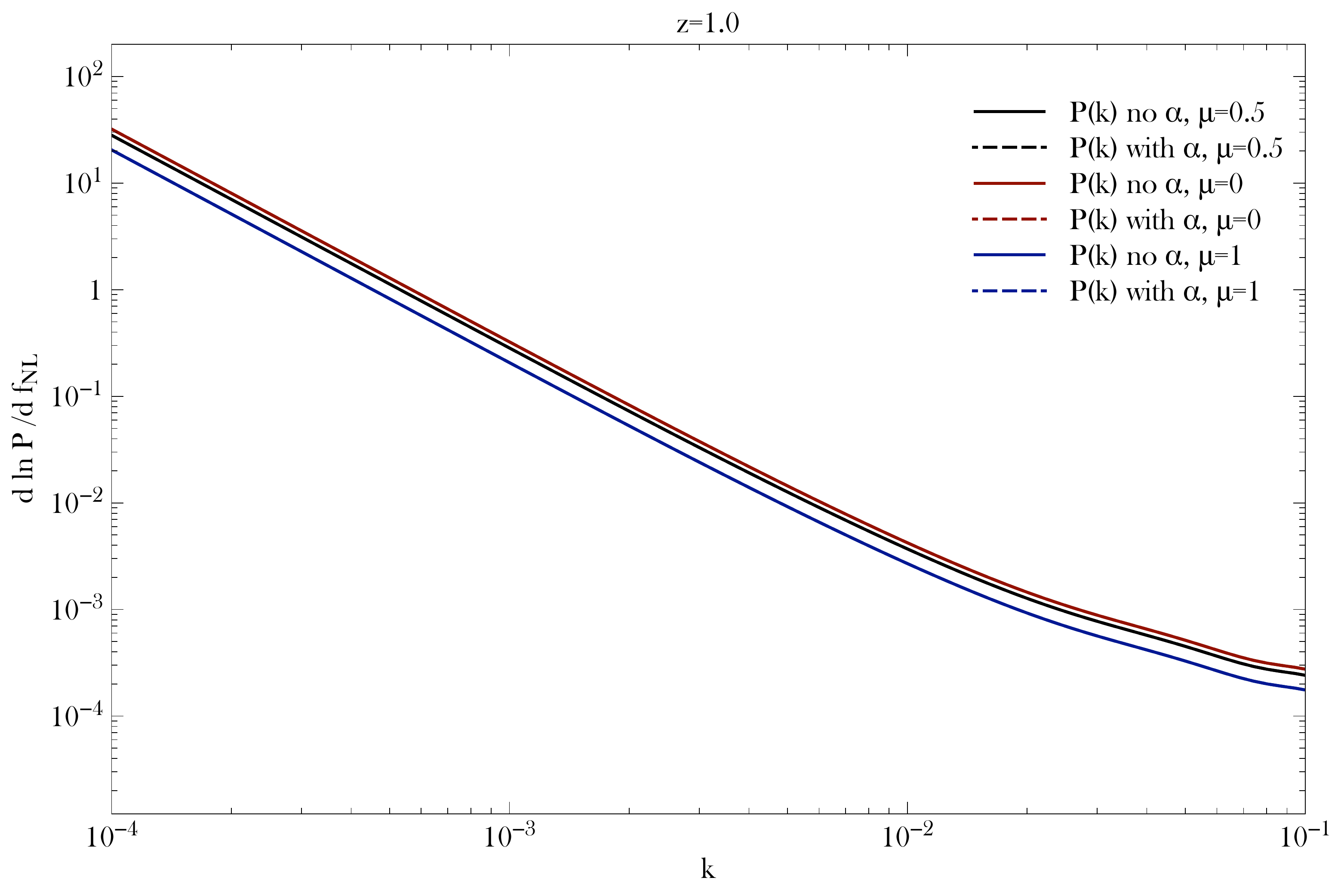}
\includegraphics[width=0.49\columnwidth]{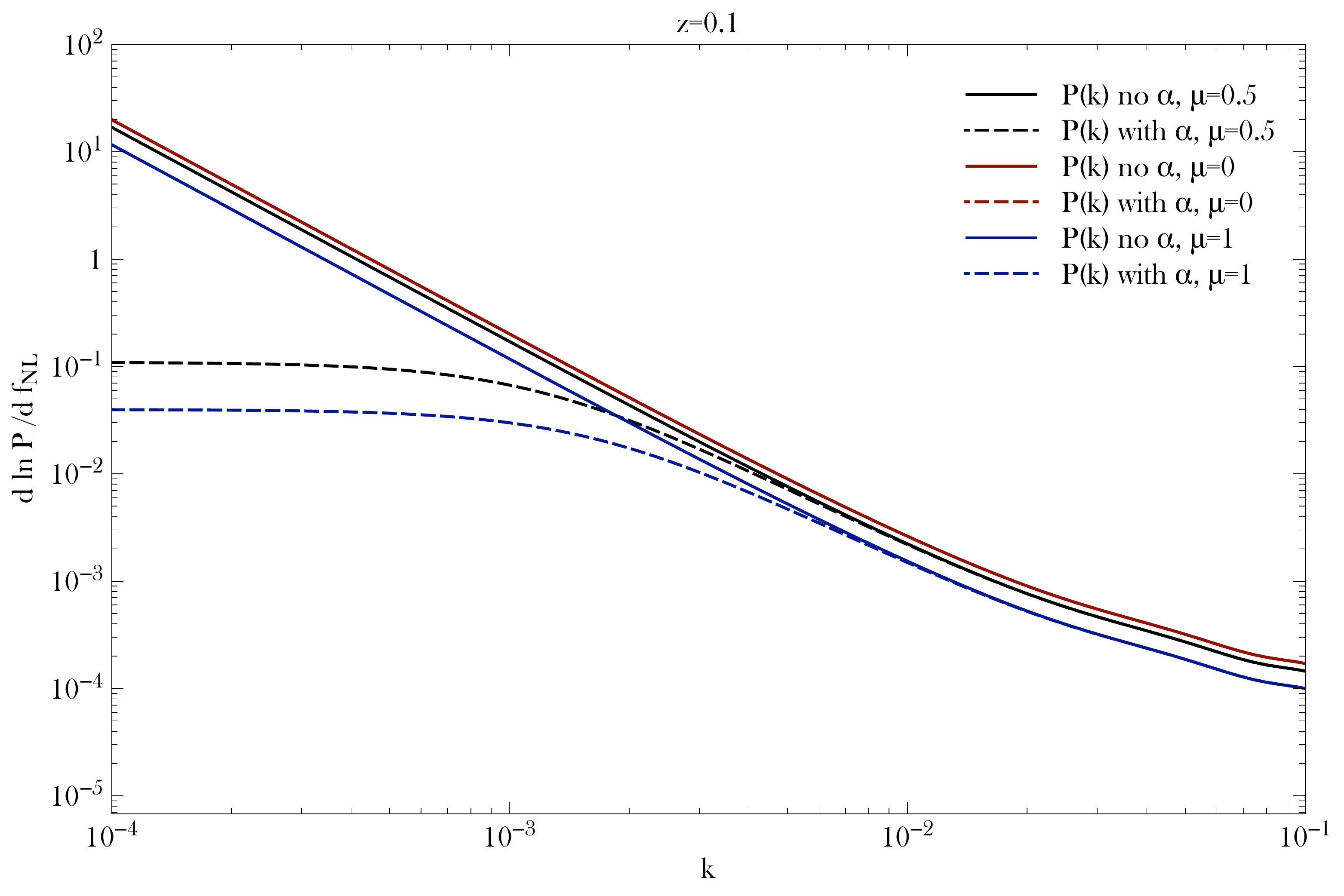}
\includegraphics[width=0.49\columnwidth]{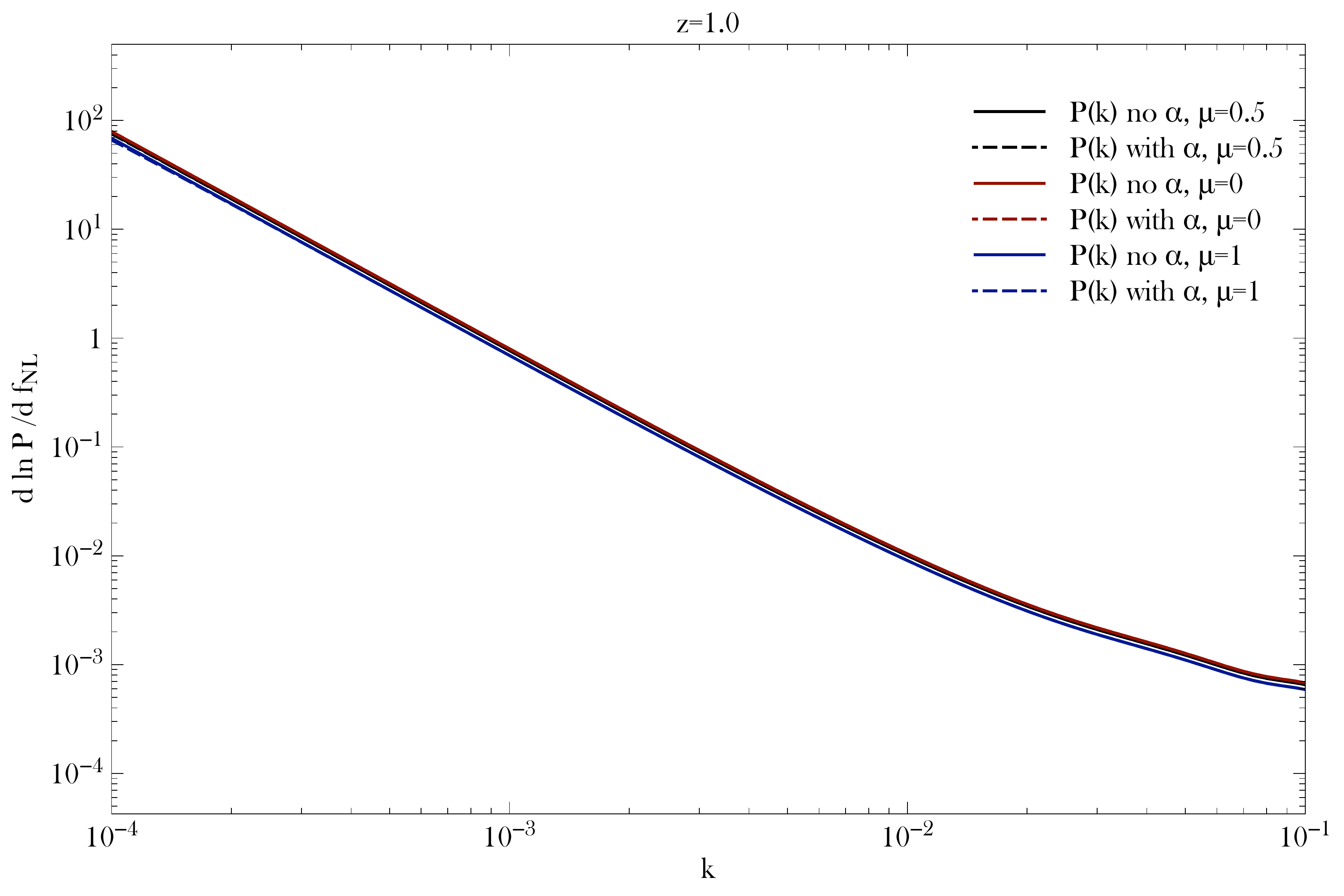}
\caption{Derivative of the power spectrum w.r.t. $f_{\rm NL}$ when including (or not) $\text{\textphnc{\Ashin}}_\alpha$, for different values of $\mu$, for $z=0.1$ (left panels) and $z=1.0$ (right panels), for the SKA (top panels) and SPHEREx (bottom panels).}
\label{fig:dPdfNL}
\end{figure*}
\end{center}

The additional contribution to the RSD operator, $\text{\textphnc{\Ashin}}_\alpha$, introduces large-scale modifications to the galaxy power spectrum (through the $k^{-2}$ dependence), so it will be somewhat degenerate with local primordial non-Gaussianity effects.
Hence, it makes sense to investigate the impact of these corrections on parameter estimation. We focus for now on measurements of the primordial non-Gaussianity parameter $f_{\rm NL}$, considering the contribution of local non Gaussianities in the squeezed limit to the rescaling of the galaxy bias~\cite{Matarrese:2000, Dalal:2008};
a more general study of how these terms affect the measurements of other parameters is left as a future work.
In Figure~\ref{fig:dPdfNL} we show the change, when including $\text{\textphnc{\Ashin}}_\alpha$, in the derivative of the (log of the) power spectrum w.r.t. $f_{\rm NL}$, for $z=0.1$ (left panels) and $z=1.0$ (right panels), for SKA (top panels) and SPHEREx (bottom panels), for different values of  the pair orientation angle $\mu$. As expected, at high redshift and for pairs perpendicular to the line of sight, Doppler terms are not affecting the derivative of the power spectrum, but for low-z and radial pairs, these effects are large.
When performing a Fisher analysis on the $0<z<0.2$ bin, we find that the difference in $\sigma(f_{\rm NL})$ when including (or not) $\text{\textphnc{\Ashin}}_\alpha$ terms can be between 10 and 100\%, depending on the survey, and most of all, $\mathcal{Q}$. This represents a very large correction; however, the full constraining power of a survey comes of course from the combination of all redshift bins, and the effects of $f_{\rm NL}$ on the power spectrum are larger on higher-$z$~\cite{Raccanelli:2014fNL}, so Doppler terms effects will be washed out by high redshift constraints.
Nonetheless, considering the precision with which future surveys aim to measure primordial non-Gaussianity (see e.g.~\cite{Alvarez:2014}), it will be wise to take into account all effects that might affect the final result.

\subsection{Angular correlations: harmonic space approach}
\label{sec:cls}
The galaxy power spectrum can also be computed in spherical harmonic space, by using the angular power spectra $C_{\ell}(z_i,z_j)$.
Here we discuss how mode-coupling and Doppler lensing terms affect measurements of angular spectra.
The geometry modifications of the Doppler term, neglected in the Kaiser formalism, are automatically included in this formalism, in which we do not perform the flat-sky approximation. For this reason, the estimate of the detectability of the Doppler effects using angular correlations $C_\ell$ is conservative and the impact of neglecting these terms in configuration space will be larger.

In this work we use a modified version of the CLASS~\cite{DiDio:2013CLASS} code to compute the galaxy correlation function, using different future galaxy redshift surveys, and we perform a $\chi^2$ analysis in order to quickly quantify the detectability of the Doppler term described in Section~\ref{sec:alpha}.
We then investigate survey requirements to reach specific larger values of the signal-to-noise ratio (SNR) for such detection; it is worth noting that a high SNR in the detection of those terms could be used to measure some cosmological parameters, and in particular it can be a direct measure of the derivative of the Hubble parameter.

We write the density, RSD and velocity terms in the angular spectra case as:
\begin{eqnarray}
\label{eq:deltas}
\Delta_{\delta}({\bf n}, z)  &=& b(z,k)\;\de^{\rm c}\left[\chi(z)\bn,\tau(z)\right] \, , \nonumber \\
\Delta_{\rm rsd}({\bf n}, z)  &=& - \frac{1}{\HH(z)}\dd_z(\bV\cdot\bn) \, , \nonumber \\
\Delta_{\rm v}({\bf n}, z)  &=& \left[b_{e}(z) - \frac{\HH'}{\HH^2}-\frac{2[1-\mathcal{Q}(z)]}{r\HH} - 2\mathcal{Q}(z)\right](\bV\cdot\bn)+ \HH \left[3-b_{e}(z)\right]\De^{-1} (\nabla\cdot\bV) \, ,
\end{eqnarray}
where $\bV$ is the peculiar velocity, $\de^{\rm c}$ is the density contrast in comoving gauge and $\HH = aH$ is the conformal Hubble parameter. All quantities are evaluated at conformal time $\tau(z)$ and at position $\chi(z)\bn=[\tau_0-\tau(z)]\bn$. Here $\chi(z)$ is the conformal distance on the light cone, $\chi(z)=\tau_0-\tau(z)$. A prime indicates a derivative w.r.t. conformal time. The  term $\De^{-1}(\nabla\cdot\bV)$ is the velocity potential, from the {\it non-local} requirement that the overdensity $\de^{\rm c}$ which is multiplied with the bias factor $b(z)$ be the comoving one.
So in this case we define the galaxy spectra $C_\ell^{\delta+{\rm rsd}, \delta+{\rm rsd}}$ as the equivalent of the $\xi^{\delta \delta}$ of Equation~\eqref{eq:xi_ii}, and the total spectra as the sum of all auto- and cross- correlations, where the $C_\ell$ can be written as:
\begin{equation}
\label{eq:Cls}
C_{\ell}^{ij} = 4\pi \int \frac{dk}{k} \Delta_{\ell}^{W_i}(k) \Delta_{\ell}^{W_j}(k) \mathcal{P}_{\mathcal{R}}(k) \;,
\end{equation}
with $\mathcal{P}_{\mathcal{R}}(k)$ being the primordial power spectrum of curvature perturbations.

We now compute our results for the spectroscopic surveys with the planned Euclid satellite~\cite{euclid}, the Square Kilometre Array~\cite{ska} (in particular the HI SKA2 galaxy survey~\cite{SKA:Abdalla}) and the proposed SPHEREx satellite~\cite{spherex}.
In the previous Section we computed the errors in the measurement of Doppler terms using the plane-parallel power spectrum; using the power spectrum allowed us to include the so-called ``mode-coupling'' terms, and Fisher analyses in Fourier space are easily computed. However, as discussed in Section~\ref{sec:fs_wa}, assuming the plane-parallel approximation excludes the wide-angle effects and (artificially) reduces the effects of velocity terms. For this reason, we repeat a detectability analysis using angular correlations, in order to include the wide angle geometry (automatically included in $C_\ell$ analyses).

We divide the galaxy redshift distributions in top-hat bins with $\Delta z = 0.04$. Using narrow redshift bins in the radial direction allows us to retain the radial information, where the velocity terms are important.
This gives us 50 bins in the range $0<z<2$ for the SKA, 25 bins in the range $0.8<z<1.8$ for Euclid, and 25 bins in the range $0<z<1$ for SPHEREx. We use the specifications for galaxy redshift distribution $N(z)$ and bias $b(z)$, and compute the evolution bias $b_{e}$ and magnification bias $\mathcal{Q}(z)$ following~\cite{Raccanelli:2015GR, Montanari:2015, spherex}.

\begin{center}
\begin{figure*}[htb!]
\includegraphics[width=0.49\columnwidth]{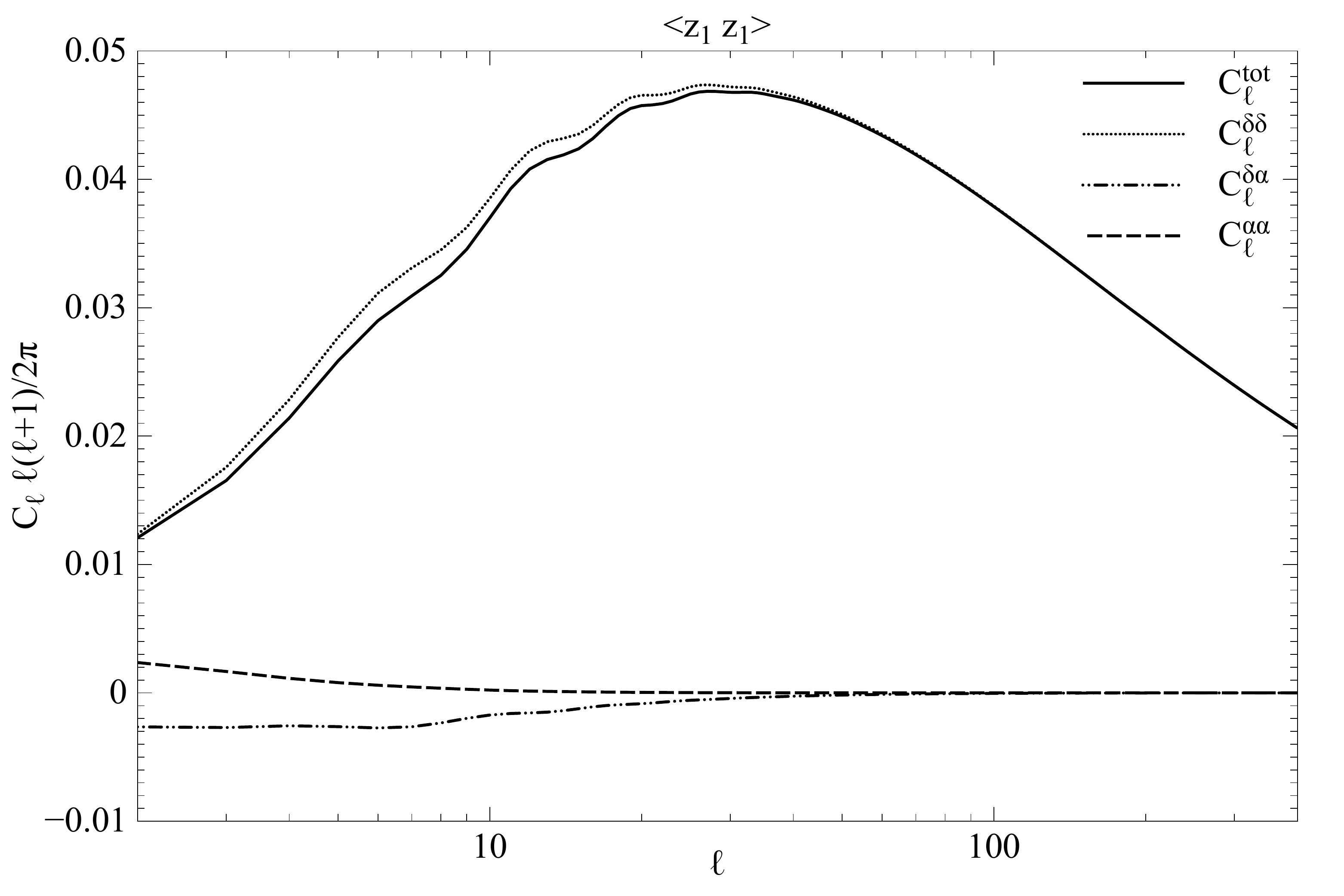}
\includegraphics[width=0.49\columnwidth]{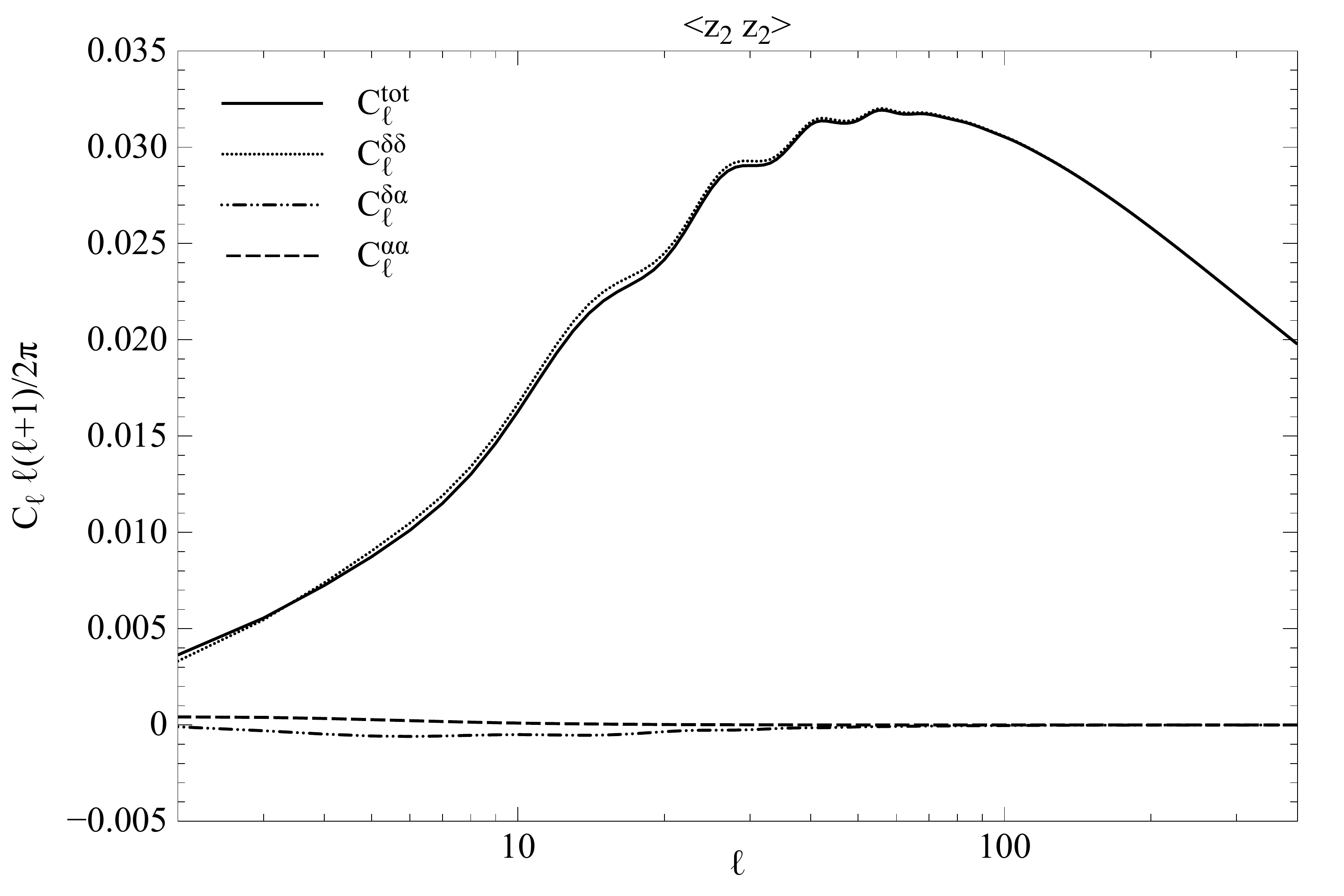}
\includegraphics[width=0.49\columnwidth]{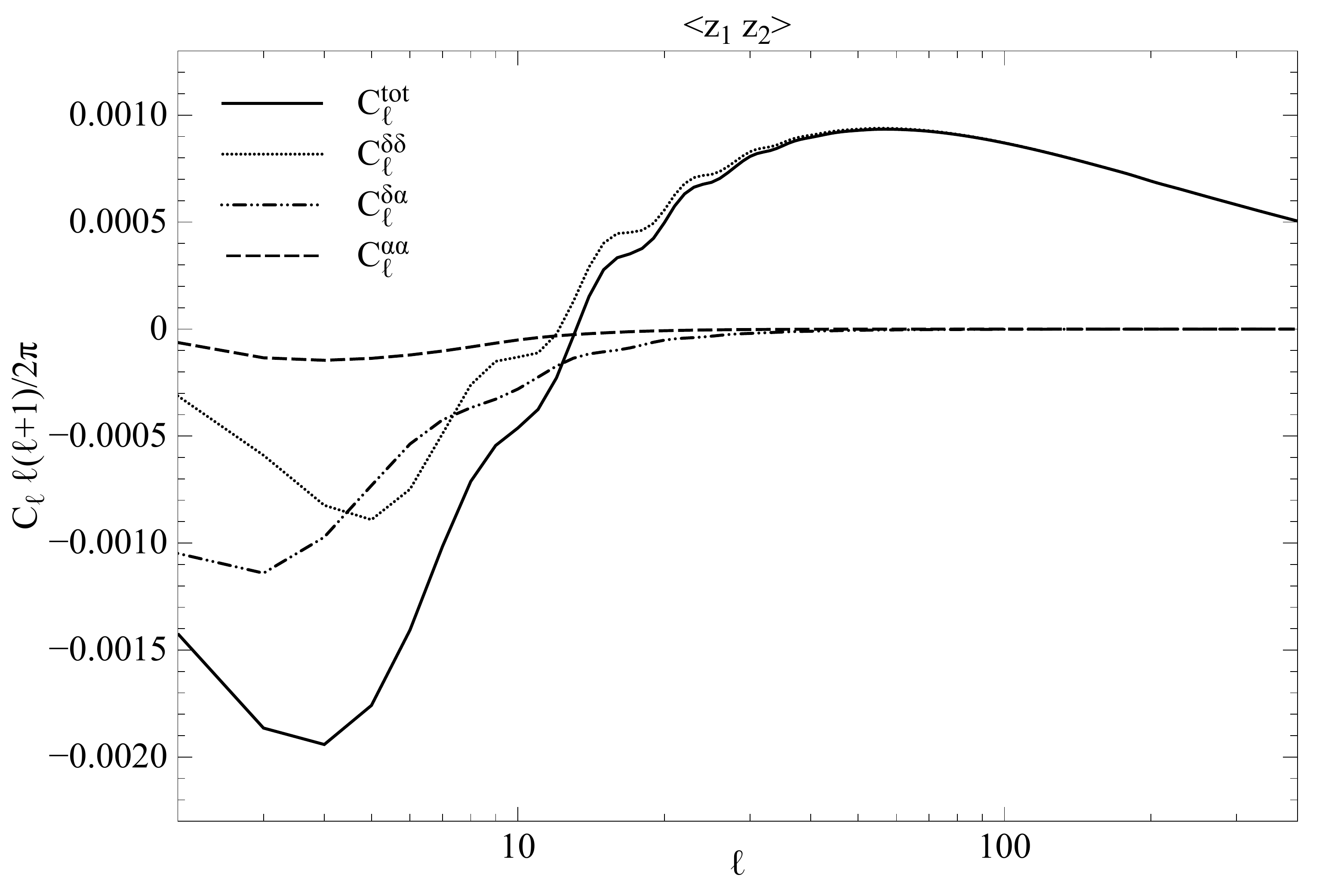}
\includegraphics[width=0.49\columnwidth]{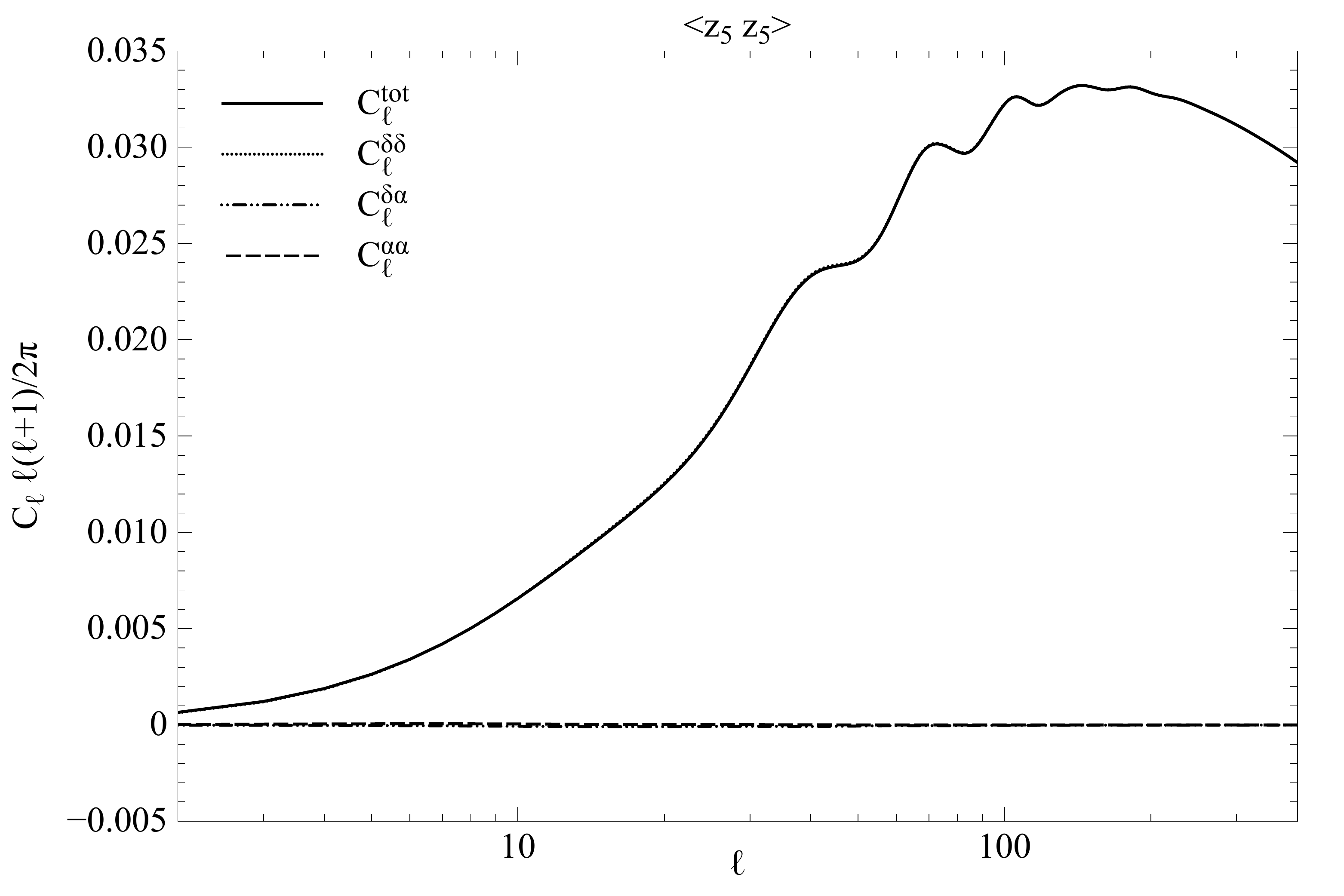}
\caption{$C_{\ell}^{XY}(z_i,z_j)$, with $\{i,j\}=\{1,1\}, \{2,2\}, \{5,5\}, \{1,2\}$ and  $\{X,Y\}=\{\delta, \alpha\}$, clockwise, assuming the SKA specifications. The number of the bin refers to the 5 sub-bins with half-width $\delta z = 0.02$ of the bin centered at $z=0.1$.
}
\label{fig:cls}
\end{figure*}
\end{center}

In Figure~\ref{fig:cls} we plot the different contributions to the total angular correlations $C_{\ell}(z_i,z_j)$ for different combinations of $(z_i,z_j)$, where indices indicate the 5 sub-bins of the bin centered at $z=0.1$.
We can see how Doppler terms can be important in radial correlations, that are otherwise negligible when using the Kaiser formalism (for an investigation of the long radial modes when including relativistic and lensing contributions, see~\cite{Raccanelli:radial, Raccanelli:2015GR}).

Then we investigate what will the signal-to-noise ratio be as a function of generic survey specifications such as number of sources detected and area of the sky surveyed, in order to understand what is the optimal configuration for the detection of Doppler effects.

We compute the signal-to-noise ratio for the detection of Doppler terms as:
\begin{equation}
\left(\frac{S}{N}\right)^2 = \nsum_{\,\,\, \ell,i,j} \frac{\left[ C_\ell^{tot}(z_i,z_j) - C_\ell^{\delta\delta}(z_i,z_j) \right]^2}{\sigma^2_{C_\ell^{tot}(z_i,z_j)}} \, ,
\end{equation}

where $C_\ell^{\delta\delta}$ here include $\Delta_{\rm rsd}$ contribution of Equation~\eqref{eq:deltas} but not the $\Delta_{\rm \textbf{v}}$ one, and the covariance is given by:
\begin{equation}
\label{eq:err-clgt}
\sigma^2_{C_{\ell \, \rm[(ij), (pq)]}} = \frac{\tilde{C}_{\ell}^{\rm (ip)} \tilde{C}_{\ell}^{\rm (jq)} + \tilde{C}_{\ell}^{\rm (iq)} \tilde{C}_{\ell}^{\rm (jp)}}{(2\ell+1)f_{\rm sky}} \, ;
\end{equation}
here $\tilde{C}_{\ell} $ denote the observed correlation multipoles including shot noise errors:
\begin{equation}
\tilde{C}_{\ell}^{ij}  = C_{\ell}^{ij} + \frac{\delta_{ij}}{dN(z_i)/d\Omega} \, ,
\end{equation}
and $dN(z_i)/d\Omega$ is the average number of sources per steradian within the bin $z_i$.
Note that we sum over the matrix indices $(i\,j)$ with $i\leq j$ and $(p\,q)$ with
$p \leq q$ which run from 1 to the number of bins.
We sum over multipoles $\ell$ limiting ourselves to linear scales, so we choose $\ell_{\rm max}(z)$ to match scales of $k_{\rm max}(z=0)\approx 0.12$.
We divide our catalog into large z bins containing 5 narrow sub-bins, and compute cross-bin correlations only within the large bins. By excluding long radial cross-correlation (i.e. for bins separated by more than 5 sub-bins), we minimize the effect of cosmic magnification effects coming from relativistic corrections along the line of sight.
In the $C_{\ell}$ for the covariance matrix we use $C_\ell^{tot}$, and we checked that gravitational potential terms represent only a small contamination to the signal (in principle time-delay effects due to gravitational potential effects can give a modification similar to Doppler lensing terms, but they are an integrated term, so at low redshift they are subdominant~\cite{Raccanelli:radial, Raccanelli:2015GR}).

We then performed the same calculation for the SKA, using 5 larger bins ($\Delta z = 0.4$) in order to confirm that our results were not caused by any numerical oscillations due to the use of narrow bins; we obtain values of the SNR slightly smaller than the case with several bins, as expected, but still consistent with the main calculation.

\begin{center}
\begin{figure*}[htb!]
\includegraphics[width=0.49\columnwidth]{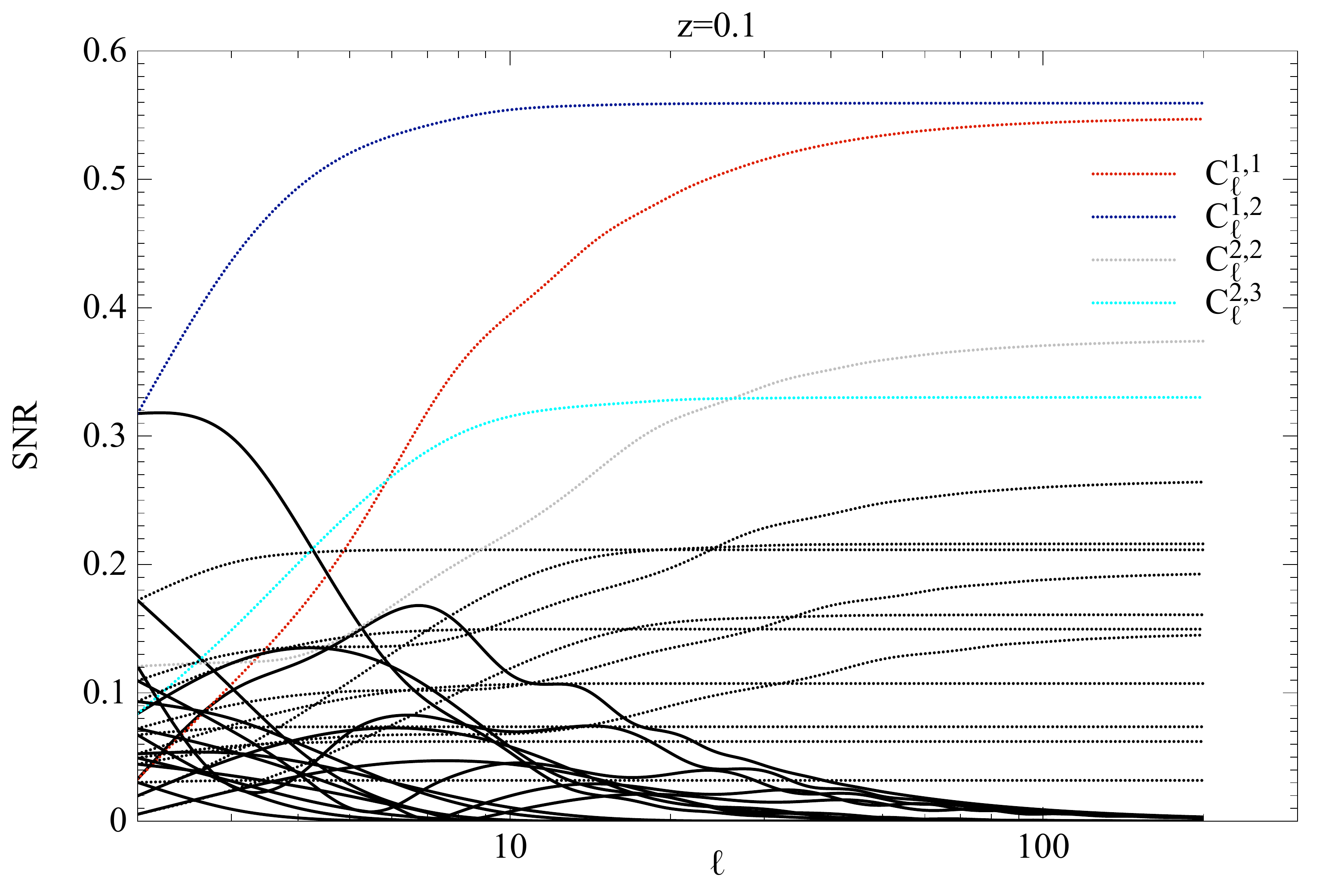}
\includegraphics[width=0.49\columnwidth]{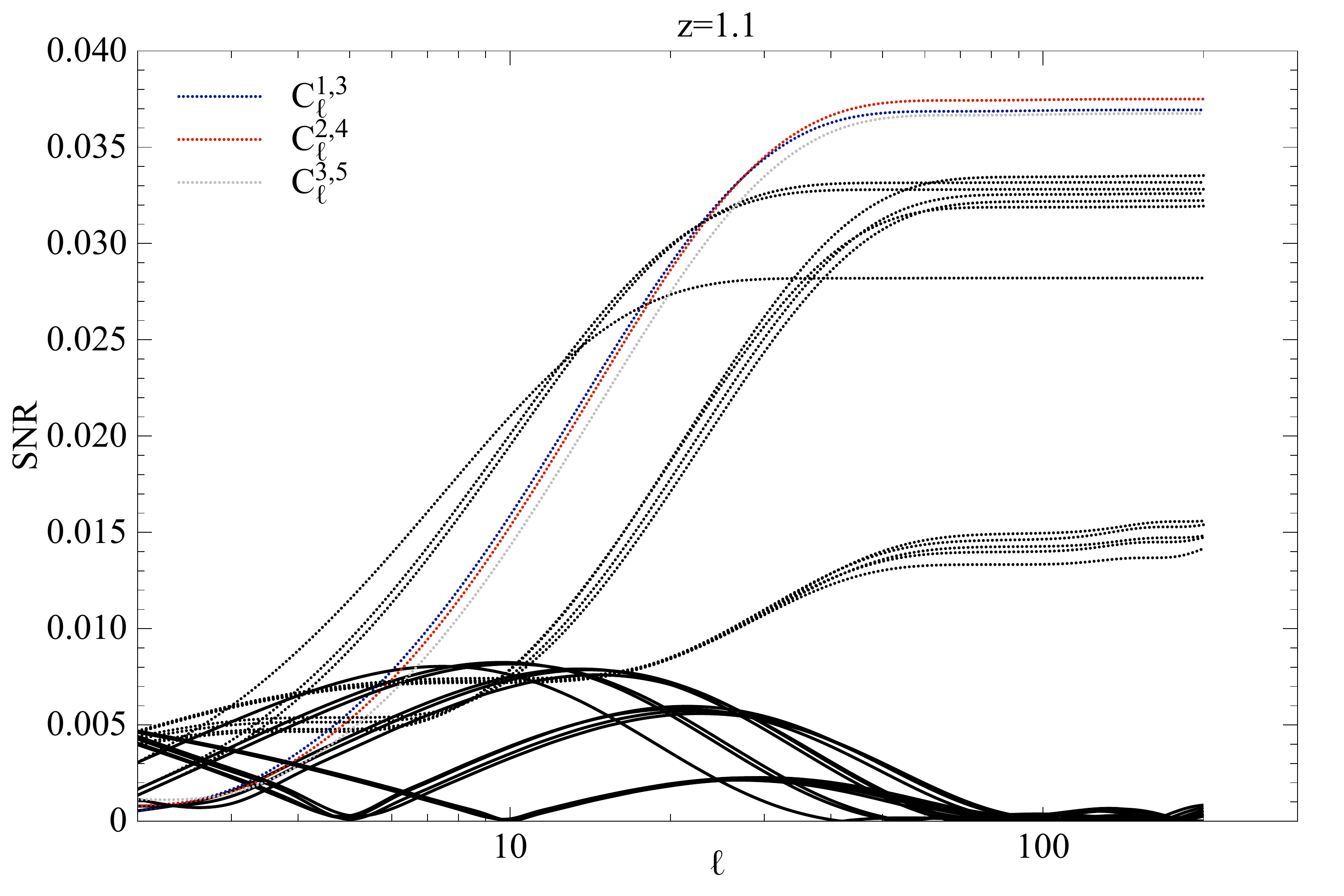}
\caption{SNR for the different correlations of sub-bins within a redshift bin centered at $z=0.1$ (left panel) and a bin centered at $z=1.1$ (right panel). Solid lines show SNR per multipole $\ell$, while dotted lines indicate the cumulative SNR. Colored are the dominant contributions.}
\label{fig:snr_comp}
\end{figure*}
\end{center}

\begin{center}
\begin{figure*}[htb!]
\includegraphics[width=0.49\columnwidth]{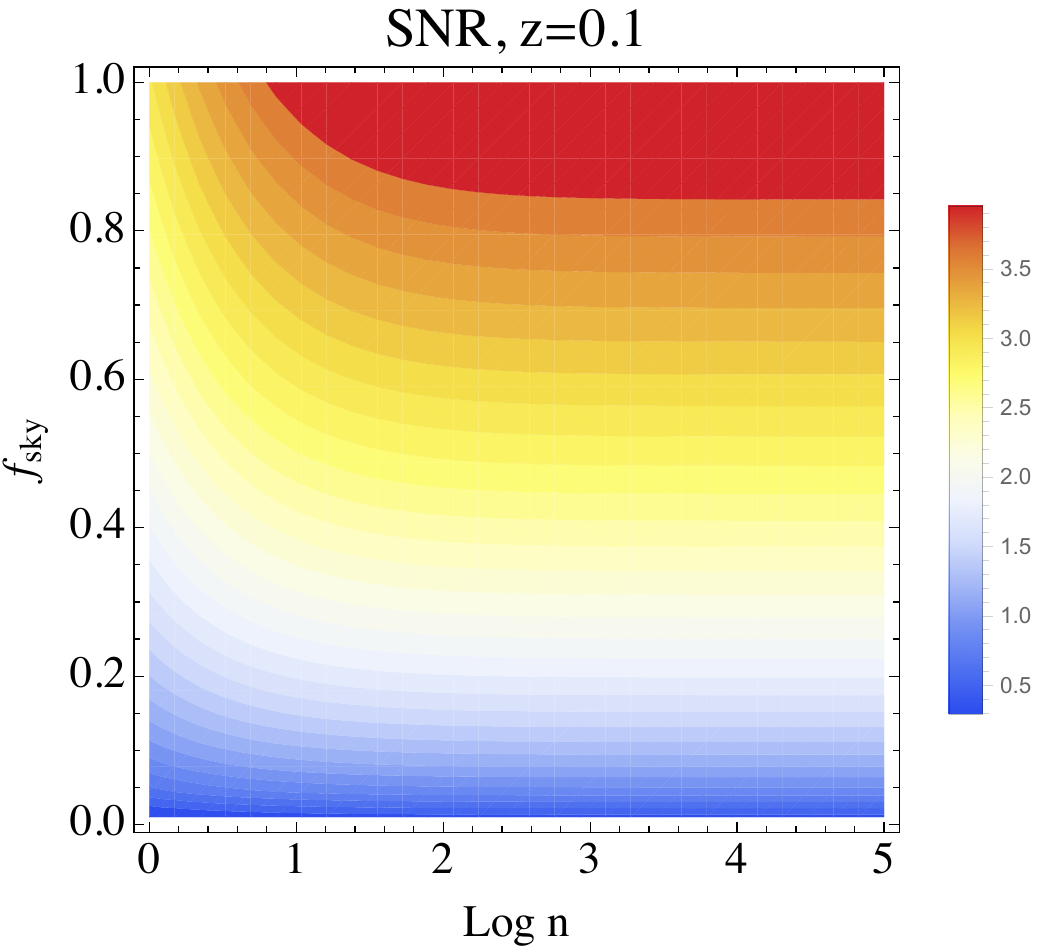}
\includegraphics[width=0.49\columnwidth]{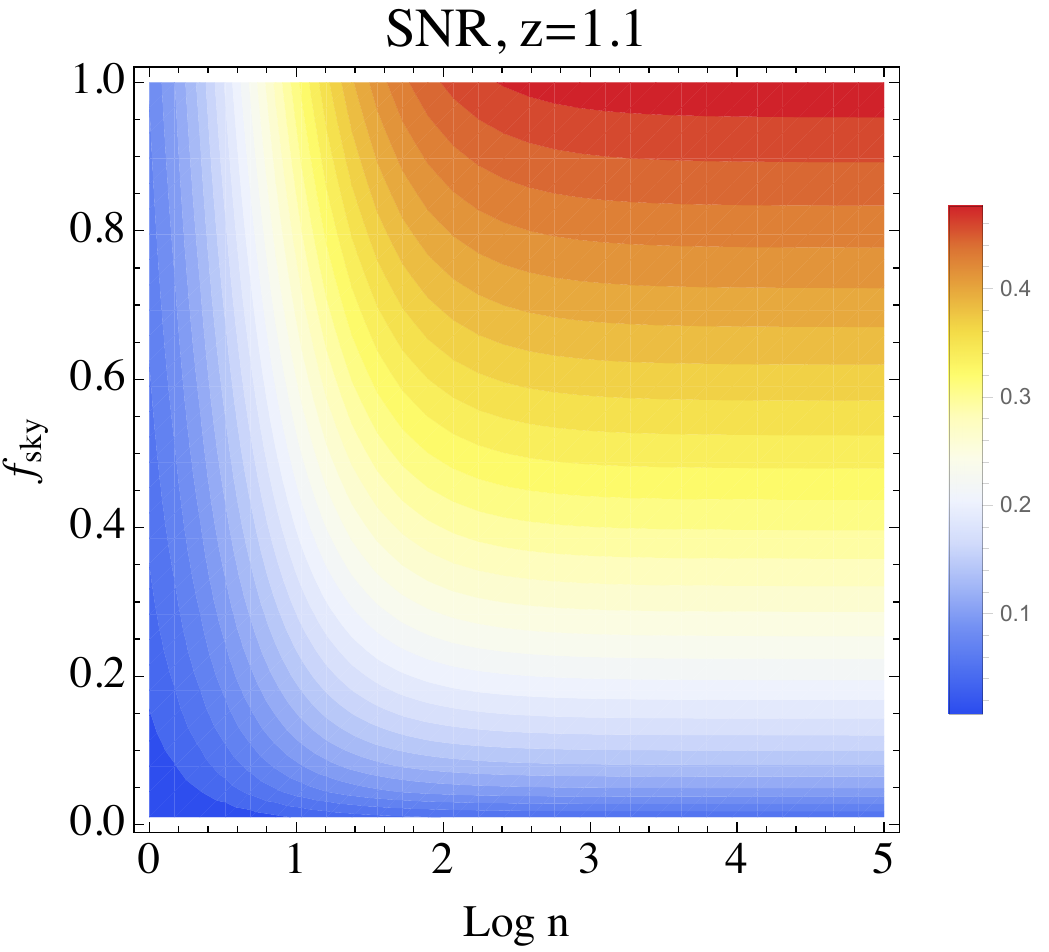}
\caption{SNR, for an SKA-{\it like} survey, for a low ($z=0.1$) and a high ($z=1.1$) redshift bins, as a function of number of sources per square degree $N$ and fraction of the sky surveyed $f_{\rm sky}$.}
\label{fig:snr_cont}
\end{figure*}
\end{center}

In Figure~\ref{fig:snr_comp} we plot the SNR for the different correlations $C_\ell(z_i,z_j)$ within a low- and a high-z redshift, for the SKA HI galaxy survey. We can see how, as expected, the SNR is much larger for the low redshift bin. Within the $z=0.1$ bin, the dominant contributions are the lowest $z$ correlations. As for the high-z bin, the largest effects are visible for radial correlations. Again this is expected, because at high-z the geometry effects are suppressed, and the Doppler lensing is mostly visible on radial correlations.
In Figure~\ref{fig:snr_cont} we plot the SNR in redshift bins with half width $\Delta_z=0.1$ (that we divide in 5 sub-bins), for a low- $(z=0.1)$ and high- $(z=1.1)$ redshift cases. The plots show results as a function of the (log of) number of galaxies per square degree and fraction of the sky observed, $f_{\rm sky}$.
We can see that, as expected, the low-z case has a larger SNR than the high-z one, that the number density of sources is important until it reaches around $100/{\rm deg.}^2$, and then saturates (the errors become cosmic-variance limited and shot noise is negligible); $f_{\rm sky}$, on the other hand, is very important for the detection of these effects, to confirm once again that Doppler effects are the most important at large angular separations. 

\begin{center}
\begin{figure*}[htb!]
\includegraphics[width=0.49\columnwidth]{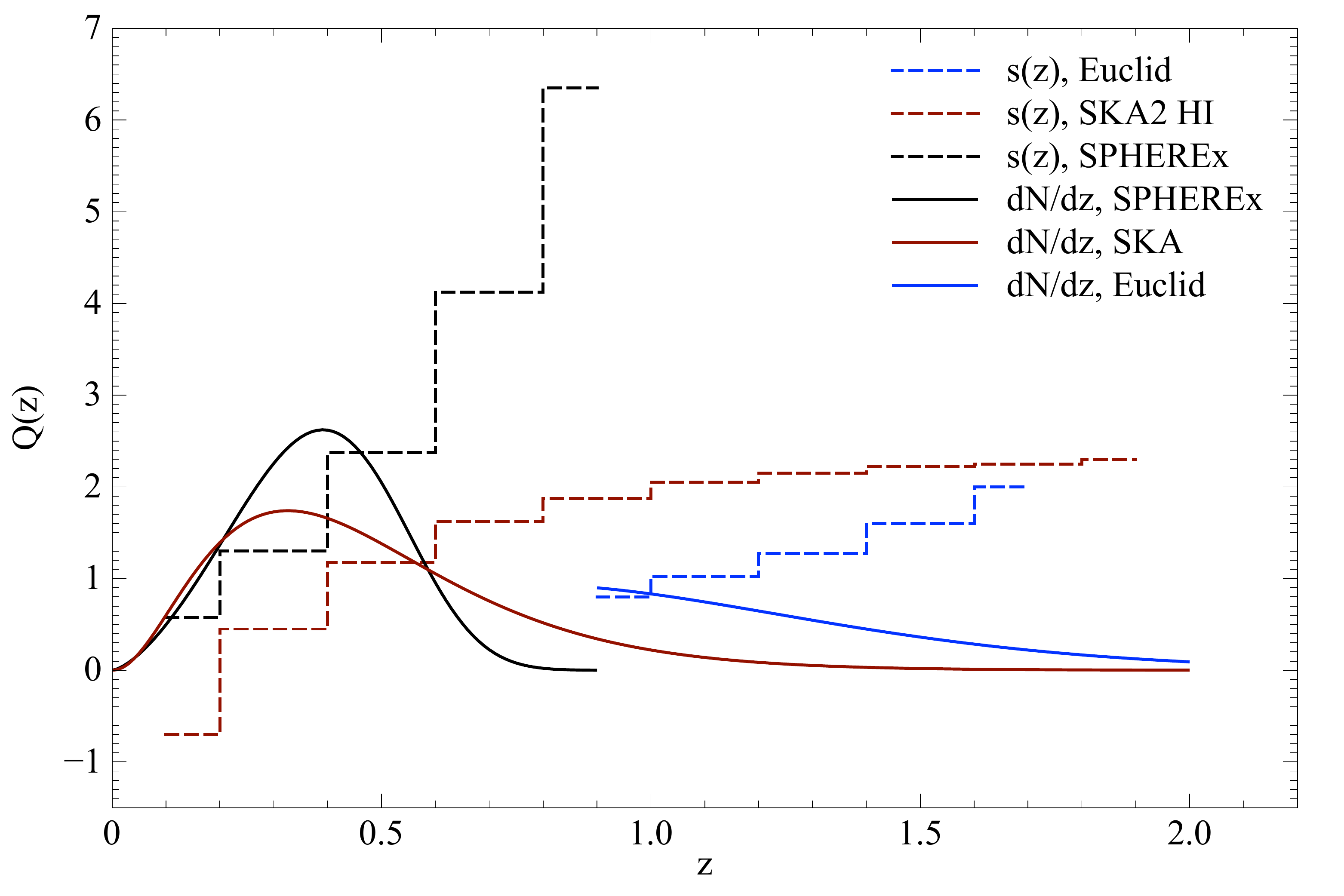}
\includegraphics[width=0.49\columnwidth]{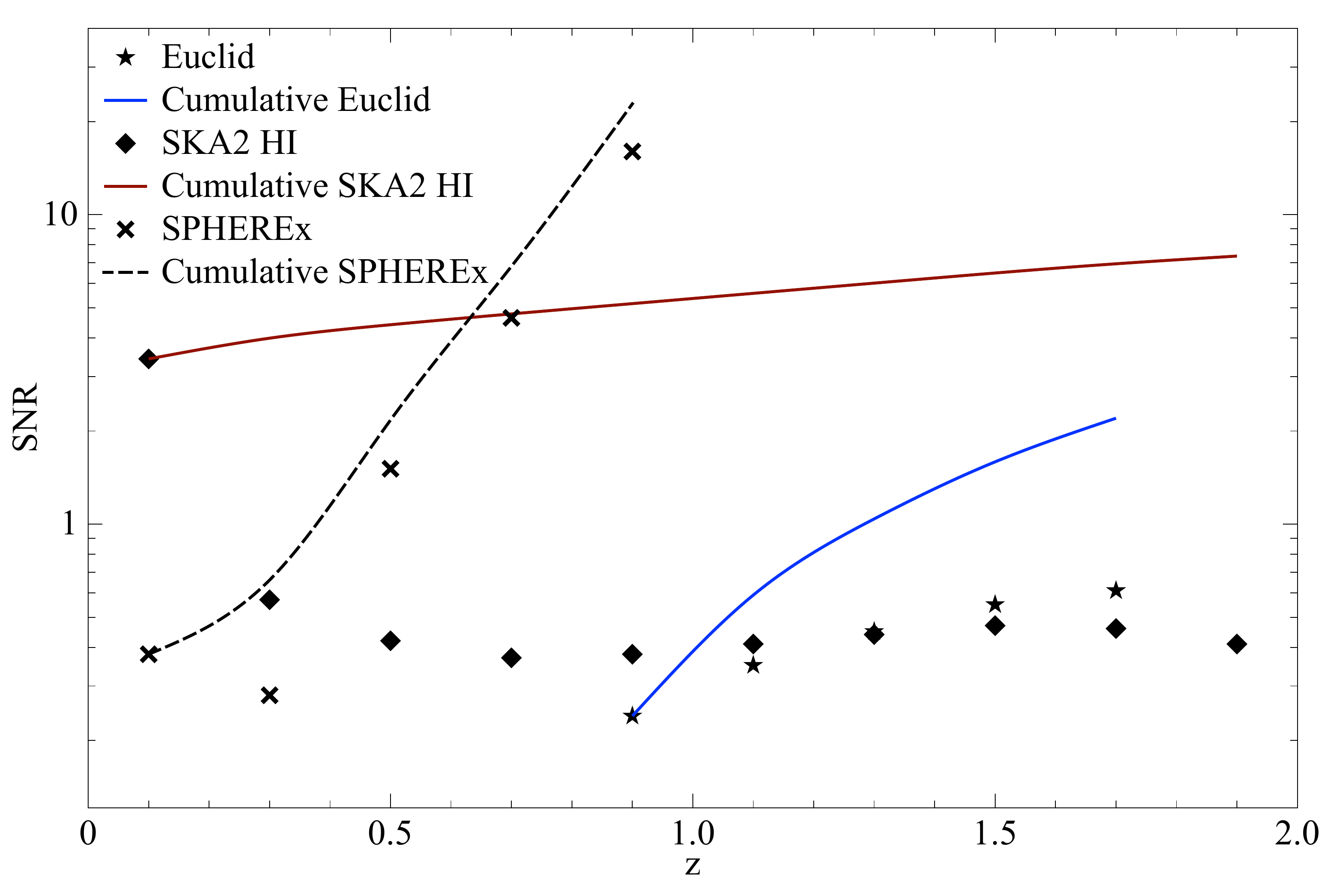}
\caption{
{\it Left Panel}: Normalized galaxy redshift distributions (solid lines) and magnification bias (dashed lines) used for the different surveys considered.
{\it Right Panel}: Signal to noise ratio per redshift bin and cumulative for different future galaxy surveys.
}
\label{fig:snr}
\end{figure*}
\end{center}

In Figure~\ref{fig:snr} we show, on the left panel, the galaxy redshift distributions and magnification bias $\mathcal{Q}(z)$ used for this work. We can see how SPHEREx has both a steeper $N(z)$ and a larger magnification bias, and this amplifies the first two terms of the $\alpha$ of~\eqref{eq:alpha_i}. On the other end, the smoother $N(z)$ and the smaller $\mathcal{Q}$ of Euclid contribute to a smaller values for $\alpha$.

On the right panel we plot the resulting SNR for Euclid, SPHEREx and the SKA.
We plot the total SNR for each one of the large bins, each of them being comprised by 5 sub-bins for which we also compute cross-bin correlations (we neglect cross-bin correlations between large bins and between sub-bins from different large bins).
The SKA and SPHEREx will both have a large number of sources at low-z and over the full sky, so they are ideally structured to detect the contribution of velocity terms. However, the differences in the slope of $N(z)$ and in $\mathcal{Q}(z)$ result in a larger SNR for SPHEREx ($\approx 18$), with most of the signal coming from intermediate redshifts, where $b_e$ and $\mathcal{Q}(z) >>1$, while most of the signal for the SKA will come from the lowest redshift bin.

It is worth noting that we assumed the standard specifications and configurations of the surveys we used as examples. However, selecting a subsample of objects with a steep selection function and a large magnification bias can amplify the signal and improve the detectability of Doppler terms. For example, a large magnitude limit results in a lower value of $\mathcal{Q}(z)$; of course one could select a subsample of objects from e.g. Euclid with a lower magnitude limit and hence obtain a very large magnification bias, resulting in a large contribution of Doppler lensing to the observed, total galaxy correlation function.

We performed a series of additional checks including the use of gaussian instead of top-hat bins, a progressive reduction in the number density and precision in the estimate of $z$ and we find that results change in a well-behaved and expected way.

The other quantity that enters in the Doppler terms is the magnification bias. In the formalism of Equation~\eqref{eq:deltas}, a value of $\mathcal{Q}=1$ suppresses lensing effects. Results will be of course dependent of its value, in the same way standard RSD results depend on the galaxy bias. A careful treatment of the modeling of $\mathcal{Q}$ will be needed in real data analyses.

This also opens up the possibility to select a subsample of galaxies with values of $\mathcal{Q}$ that are as far away as possible from 1, in order to increase the SNR of the Doppler terms; we will investigate this in our follow-up paper.

\section{Discussion and conclusions}
\label{sec:conclusions}
In this paper we investigated the importance of Doppler terms in the galaxy correlation function and their potential detectability using future galaxy surveys.

The Doppler terms in the galaxy correlation function in configuration (redshift) space can be divided into three different effects: geometry/velocity corrections, Doppler lensing and cosmic acceleration effects.

The {\it geometry/velocity} terms account for the fact that galaxy pairs move from low-density to high-density regions; it's a Doppler effect due to local infall velocity toward overdense regions.
This effect exists in also in the flat-sky approximation, but it is very small when $\theta \rightarrow 0$. This can be seen mathematically because, looking at Equation~\eqref{eq:xi_ii} (and Equations 4,5 of~\cite{Papai:2008}), the \aaa terms are suppressed by a factor that depends on $sin(\theta)$, so for small $\theta$ the \aaa terms are small.
Physically, this can be understood as the fact that the presence of an observer and the local motion of galaxies along the (non-parallel) line of sight enhance the differences between {\bf r} and {\bf s} and modifies the value of the angle with the line of sight $\mu$.
That is the reason why these effects are called mode-coupling, because they cause information to ``leak'' from the diagonal to off-diagonal parts of the RSD operator.
This was first noticed to be a potentially important effect in~\cite{Papai:2008} and proven with simulations~\cite{Raccanelli:2010}, where the mode-coupling effects were shown to be necessary to fit simulated data (see Figure 2 and 4 of~\cite{Raccanelli:2010}, where the second one also recovers the results of the right panel of Figure 1 of~\cite{Papai:2008}).
In~\cite{Samushia:2012} these effects were taken into consideration for SDSS-II DR7 data, and found that they were not important. However, at sufficiently large scales, their amplitude was on the order $\mathcal{O}(10^{-1})$ compared to statistical errors. Considering the increment in area surveyed and number of sources observed by future surveys, it is apparent how these effects will need to be taken into account for future galaxy clustering analyses.

The {\it Doppler lensing} terms account for apparent modifications of angular size and magnitude of a galaxy, due to the motion of the source with respect to the observer.
Galaxies with the same (apparent) redshift space position can be, in real space, more distant (if they are moving toward us, $V \cdot \hat{n} < 0$), than galaxies with no peculiar velocity, or closer to us (if $V \cdot \hat{n} > 0$). So in practicethree galaxies can be seen at the same (apparent) position, in $z$-space, while in reality, they are at three different distances compared to their apparent position: one closer (if receding), one correct (if stationary), and one farther (if approaching).
This generates two opposite effects on the observed angular size of the galaxy: a galaxy that is more distant in real space is observed under a smaller solid angle, but its photons were emitted at an earlier time, when the value of the scale factor of the universe was smaller.
Hence, those photons experienced a larger stretch in their path toward us, which increases the observed angular size of the galaxy.
The overall effect depends on z and {\bf v}.
These effects were partially investigated in the context of studying relativistic effects in e.g.~\cite{Bonvin:2014}. The same Doppler lensing but in the context of lensing analyses was recently investigated in~\cite{Bacon:2014}, where using simulations it was found that these terms can be used for cosmology with lensing measurements and they even dominate over magnification effects at low z.

{\it Cosmic acceleration} translates into volume distortions. The fact that we observe galaxies with velocity terms along the past lightcone causes in practice a modification of the volume of the cosmological volume surveyed.

In our study we first wrote the 2-point galaxy correlation function using a formalism that separates contributions from the density, doppler, and mixed terms, and showed their impact on the 2D galaxy correlation function.
We find that, as expected, the Doppler effects are negligible at small angular separation and high-z; however, when dropping the flat-sky approximation and using a proper modeling of wide-angle correlations, Doppler terms can become relevant.
After describing the general form of the \aaa terms that models these effects and noticing that it depends on a variety of physical effects and cosmological parameters, including the time derivative of the Hubble factor, we investigate their potential detectability.
We compute redshift-dependent angular correlations for the planned Euclid, SPHEREx and SKA surveys, and confirmed that these effects could be detected by a survey with a large number of sources observed over a wide part of the sky at low redshift.

We then showed that the effect of Doppler terms in Fourier space has a $k^{-2}$ scale-dependence, which can therefore mimic the effects of primordial non-Gaussianity. We quantified the effect to be inducing an $f_{\rm NL}^{\rm eff}$ parameter of a few (the exact value being dependent on the value of magnification bias $\mathcal{Q}$).
This would consequently affect the estimate of the precision in measuring $f_{\rm NL}$ for future surveys. We investigated this effect and found that neglecting Doppler terms in the low-z bins would cause a large error in the forecasts of $\sigma (f_{\rm NL})$.

Doppler terms have been considered in the past, and were found to be not important for past and current surveys. In this work we show how the signal exists in a specific part of the $\{z, f_{\rm sky}, n_g, \mathcal{Q}\}$ parameter space, and is suppressed when assuming the flat-sky approximation and when averaging over $z$. Detecting a large number of sources with large magnification bias at low-z over a large fraction of the sky, and performing a tomographical analysis would allow the possible detection of Doppler effects using galaxy redshift surveys.

We consider a few possible uncertainties that could affect measurements of Doppler terms, and found that the most important factors for a reliable measurement are a good knowledge of the redshift distribution and the magnification bias of the observed galaxies. However, the importance of magnification bias opens up the possibility of enhancing the signal we are looking for, by selecting sources with a large value of $\mathcal{Q}$, in the same way highly biased objects are used in standard RSD analyses; it is worth considering also the possibility of extending the multi-tracer technique~\cite{Mcdonald:2009} to populations with different magnification bias.
The general conclusions of this paper can be used as a guide to understand what are the best configurations for detecting Doppler effects (or to minimize them, if they are believed to be a contaminant). Our results can allow the selection of sub-samples of galaxies in a way that enhances (or minimizes) the mode-coupling or the Doppler lensing terms, opening up the possibility of measuring new effects using galaxy clustering.

We will investigate all these subtleties and look for a possible detection of Doppler terms in a more data oriented paper in preparation.

\vspace{0.5 cm}

{\bf Acknowledgments:}\\
We thank Francesco Montanari, Joe Silk, Istv\'{a}n Szapudi and Roland de Putter for useful contributions, Enea Di Dio and David Alonso for helpful discussions.
AR is supported by the John Templeton Foundation.
During the preparation of this work DB was supported by the Deutsche Forschungsgemeinschaft through the Transregio 33, The Dark Universe. 
DB acknowledges the hospitality of the Department of Physics \& Astronomy, Johns Hopkins University, where part of this work was carried out.

\appendix
\section{Derivation of Equations~\eqref{eq:xi_ii}}
\label{app:xi}
In this Appendix we illustrate the derivation of Equations~\eqref{eq:xi_ii}, where we write the galaxy two-point correlation function separating terms including density and RSD (in wide-angle), pure Doppler term, and mixed terms.
Starting from~\cite{Szalay:1997, Papai:2008}, we can re-arrange terms and express them in terms of the double angles. We have:
\begin{align}
\xi^{\delta \delta}(s,\theta,\varphi) &= \xi_0^2\left[\frac{1}{15}\left[15+10 f + 2 f^2 + f^2 \cos 4\theta \right] \right] + \nonumber \\
& +  \xi_2^2\left[ -\frac{1}{42} f \left[ 14+4f+2f\cos 4\theta +3(7+3f)\cos 2(\theta-\varphi)+21\cos 2(\theta+\varphi) +9f\cos 2(\theta+\varphi) \right] \right] \\ 
& +  \xi_4^2\left[ \frac{1}{280} f^2\left[6+3\cos 4\theta +10\cos 2(\theta-\varphi)+35\cos 4\varphi+10\cos 2(\theta+\varphi) \right] \right] \nonumber \\ 
\xi^{\delta \alpha}(s,\theta,\varphi) &= \frac{\alpha}{s} \Bigg\{ \\ \nonumber
&\xi_1^1 \left[f \cos 2\varphi \csc(\theta - \varphi) \csc(\theta + \varphi) \sin^2 2\theta  
 + \frac{1}{5} f^2 \csc(\theta - \varphi) \csc(\theta + \varphi) \sin 2\theta  [\cos 2\varphi \sin 2\theta  + \sin 4\theta ] \right] \nonumber \\ 
& + \xi_3^1 \left[-\frac{1}{40} f^2 \right] \left[ \cos 2\theta -\cos 6\theta  - 3\cos (4\theta-2\varphi)  + 6\cos 2\varphi - 3\cos(4\theta+2\varphi) \right]\csc(\theta - \varphi) \csc(\theta + \varphi) \Bigg\} 
\nonumber \\
\xi^{\alpha \alpha}(s,\theta,\varphi) &= \frac{\alpha^2}{s^2g_1g_2} \Bigg\{ \xi_0^0 \left[ \frac{4}{3} f^2 \cos 2\theta  \right] + 
\xi_2^0 \left[ -\frac{2}{3} f^2 \left(\cos 2\theta +3\cos 2\varphi \right) \right] \Bigg\} \, .
\end{align}

First we will use the trigonometric identities
\begin{align}
	\sin(x+y) \sin(x-y) =& \frac{1}{2}[ \cos 2y - \cos 2x ]\\
	\cos(x+y)+\cos(x-y) =& 2 \cos x \cos y\\
	1+\cos 4x = 2 \cos^2 2x
\end{align}

Let us consider the $\xi^{\delta \delta}$ term. Starting with the coefficient of $ \xi_0^2$:
\begin{align}
\frac{1}{15}&\left(15+10 f + 2 f^2 + f^2 \cos(4\theta) \right)=
\frac{1}{15}\left[15+10 f +  f^2 + f^2 (1+ \cos 4\theta) \right]=
\\
\frac{1}{15}&\left[15+10 f +  f^2 +2 f^2 \cos^2 2\theta \right] \nonumber 
\end{align}

Next, let us simplify the coefficient of $\xi_2^2$:
\begin{align}
& -\frac{f}{42} \left( 14+4f+2f\cos(4\theta)+3(7+3f)\cos[2(\theta-\varphi)]+21\cos[2(\theta+\varphi)]+9f\cos[2(\theta+\varphi)]\right)=
\\
 &-\frac{f}{42} \left(14+4f+2f\cos(4\theta) + (21+9f) \left[\cos[2(\theta-\varphi)]+\cos[2(\theta+\varphi)]\right] \right)=
\nonumber\\
 &-\frac{f}{42} \left(14+2f+2f(1+\cos 4\theta) + (21+9f) \left[\cos 2\theta+\cos 2\varphi \right] \right)
\nonumber \\
 &-\frac{f}{42} \left[14+2f+4f \cos^2 2\theta + (21+9f) \left(\cos 2\theta+\cos 2\varphi \right) \right] \nonumber
\end{align}

Next, the coefficient of $ \xi_4^2$:

\begin{align}
\frac{1}{280}&
f^2\left[6+3\cos(4\theta)+10\cos[2(\theta-\varphi)]+35\cos(4\varphi)+10\cos[2(\theta+\varphi)]\right]=
\\
\frac{1}{280}&
f^2\left[6+3\cos(4\theta)+10\left[\cos[2(\theta-\varphi)]+\cos[2(\theta+\varphi)]\right]
+35\cos(4\varphi)\right]=
\nonumber\\
\frac{1}{280}&f^2\left[6+3\cos 4\theta+35\cos 4\varphi +20(\cos 2\theta+\cos 2\varphi)\right]=
\nonumber\\
\frac{f^2}{280}&\left[6+3\cos 4\theta+35\cos 4\varphi +20(\cos 2\theta+\cos 2\varphi)\right] \nonumber
\end{align}

Summarizing:

\begin{align}
\xi^{\delta \delta}(s,\theta,\varphi) = 
\xi_0^2&\left(\frac{1}{15}\right)\left[15+10 f +  f^2 +2 f^2 \cos^2 2\theta \right]+ 
\\
-\xi_2^2 &\left( \frac{f}{42}\right) \left[14+2f+4f \cos^2 2\theta + (21+9f) \left(\cos 2\theta+\cos 2\varphi \right) \right]
\nonumber\\
+\xi_4^2& \left(\frac{f^2}{280}\right) \left[6+3\cos 4\theta+35\cos 4\varphi +20(\cos 2\theta+\cos 2\varphi)\right]
\nonumber
\end{align}

After this, let us simplify the $\xi^{\delta \alpha}$ term. Both of the terms contain the same expression that can be pulled out to the front. 
\begin{align}
\xi^{\delta \alpha}
=&\Big(\frac{\alpha}{s} \big)\csc(\theta - \varphi) \csc(\theta + \varphi)  \Bigg\{
\\
&\xi_1^1 \left[f \cos(2\varphi) \sin^2(2\theta) + \frac{1}{5} f^2  \sin(2\theta) [\cos(2\varphi)\sin(2\theta) + \sin(4\theta)] \right] + 
\nonumber \\ 
+& \xi_3^1 \left[-\frac{1}{40} f^2 \left[ \cos(2\theta)-\cos(6\theta) - 3\cos(4\theta-2\varphi) + 6\cos(2\varphi) - 3\cos[2(2\theta+\varphi)] \right] \right]\Bigg\} \, \nonumber
\end{align}

We can write the two cosecants as
\begin{equation}
\csc(\theta - \varphi) \csc(\theta + \varphi)  = \frac{1}{\sin(\theta - \varphi) \sin(\theta + \varphi)}
= \frac{2}{\cos 2 \varphi - \cos 2 \theta} 
\end{equation}

\begin{align}
\xi^{\delta \alpha}
=&\frac{2 \alpha}{s(\cos 2 \varphi - \cos 2 \theta)}  \Bigg\{\xi_1^1 \left[f \cos(2\varphi) \sin^2(2\theta) + \frac{1}{5} f^2  \sin(2\theta) [\cos(2\varphi)\sin(2\theta) + \sin(4\theta)] \right]  \\ 
+& \xi_3^1 \left[-\frac{1}{40} f^2 \left[ \cos(2\theta)-\cos(6\theta) - 3\cos(4\theta-2\varphi) + 6\cos(2\varphi) - 3\cos[2(2\theta+\varphi)] \right] \right]\Bigg\} \, \nonumber
\end{align}

Now the coefficient of $\xi_1^1$:

\begin{align}
f &\cos 2\varphi \sin^2 2\theta + \frac{1}{5} f^2  \sin2\theta [\cos 2\varphi \sin 2\theta + \sin 4\theta] =
\\
f &\cos 2\varphi \sin^2 2\theta + \frac{1}{5} f^2  \sin2\theta [\cos 2\varphi\sin 2\theta + 2 \sin 2\theta \cos 2\theta] =
\nonumber \\
f & \sin^2 2\theta  \Big[ \cos(2\varphi) + \frac{f}{5}  [\cos 2\varphi  + 2 \cos 2\theta] \Big]=
\nonumber \\
f & \sin^2 2\theta  \Big[ (1+  \frac{f}{5})\cos 2\varphi +  \frac{2f}{5} \cos 2\theta \Big]\nonumber
\end{align}

And the coefficient of $ \xi_3^1$:

\begin{align}
-\frac{1}{40} f^2 &\left[ \cos 2\theta -\cos 6\theta+ 6\cos 2\varphi  - 3\cos (4\theta-2\varphi) - 3\cos(4\theta+2\varphi)] \right] =
\\
-\frac{1}{40} f^2 &\left[ \cos 2\theta -\cos 6\theta+ 6\cos 2\varphi  - 3\cos (4\theta-2\varphi) - 3\cos(4\theta+2\varphi)] \right] =\nonumber
\\
-\frac{1}{40} f^2 &\left[ -3\cos 2\theta +4\cos^3 2\theta+ 6\cos 2\varphi  - 6 \cos  4\theta \cos 2\varphi \right]= \nonumber
\\
-\frac{1}{40} f^2 &\left[ -3\cos 2\theta +4\cos^3 2\theta+ 6\cos 2\varphi (1 -  \cos  4\theta)\right] =\nonumber
\\
-\frac{1}{40} f^2 &\left[ -3\cos 2\theta +4\cos^3 2\theta+ 12 \cos 2\varphi \sin^2  2\theta \right] \nonumber
\end{align}

Summarizing:

\begin{align}
\xi^{\delta \alpha}
=\frac{2 \alpha}{s(\cos 2 \varphi - \cos 2 \theta)}&  
\Bigg\{ \xi_1^1\  f  \sin^2 2\theta  \left[ (1+  \frac{f}{5})\cos 2\varphi +  \frac{2f}{5} \cos 2\theta \right] 
\\ 
+ \xi_3^1\ &\Bigg[-\frac{ f^2}{40}\left[ -3\cos 2\theta +4\cos^3 2\theta+ 12 \cos 2\varphi \sin^2  2\theta \right] \Bigg]\Bigg\}\nonumber
\end{align}

Finally the $\xi^{\alpha \alpha}$ term is simple as is:
\begin{align}
\xi^{\alpha \alpha}(s,\theta,\varphi) &= \frac{\alpha^2}{s^2g_1g_2} \Bigg\{ \xi_0^0 \left[ \frac{4}{3} f^2 \cos 2\theta \right] + 
\xi_2^0 \left[ -\frac{2}{3} f^2 \left(\cos 2\theta +3\cos 2\varphi \right) \right] \Bigg\} \, .
\end{align}

First, we can write $g_1 g_2$ as
\begin{align}
	g_1 g_2 =&  \frac{\cos 2 \varphi - \cos 2 \theta}{\sin^2 2 \theta}
\end{align}

With this we can express
\begin{align}
\xi^{\alpha \alpha} &= \frac{\alpha^2}{s^2}  \frac{\sin^2 2 \theta}{\cos 2 \varphi - \cos 2 \theta}
 \Bigg\{ \xi_0^0 \left[ \frac{4}{3} f^2 \cos 2\theta \right] + 
\xi_2^0 \left[ -\frac{2}{3} f^2 \left(\cos 2\theta +3\cos 2\varphi\right) \right] \Bigg\} \, .
\end{align}

Copying all this into one place:
\begin{align}
\xi^{\delta \delta}(s,\theta,\varphi) =\ &\xi_0^2\left(\frac{1}{15}\right)\left[15+10 f +  f^2 +2 f^2 \cos^2 2\theta \right] 
\\
-&\xi_2^2 \left( \frac{f}{42}\right) \left[14+2f+4f \cos^2 2\theta + (21+9f) \left(\cos 2\theta+\cos 2\varphi \right) \right]
\nonumber \\
&\xi_4^2 \left(\frac{1}{280}\right) \left[6+3\cos 4\theta+35\cos 4\varphi +20(\cos 2\theta+\cos 2\varphi)\right]
\nonumber\\
\xi^{\delta \alpha}(s,\theta,\varphi)
=\frac{2 \alpha}{s(\cos 2 \varphi - \cos 2 \theta)} & \Bigg\{
\xi_1^1\  f  \sin^2 2\theta  \left[ (1+  \frac{f}{5})\cos 2\varphi +  \frac{2f}{5} \cos 2\theta \right] 
\\ 
&+\xi_3^1\ \Bigg[
	-\frac{1}{40} f^2\left[ -3\cos 2\theta +4\cos^3 2\theta+ 12 \cos 2\varphi \sin^2  2\theta \right] 
\Bigg] \Bigg\} \nonumber
\end{align}
\begin{align}
\xi^{\alpha \alpha}(s,\theta,\varphi) 
&= \frac{\alpha^2}{s^2}  \frac{\sin^2 2 \theta}{\cos 2 \varphi - \cos 2 \theta}
 \Bigg\{ \xi_0^0 \left[ \frac{4}{3} f^2 \cos 2\theta  \right] + 
\xi_2^0 \left[ -\frac{2}{3} f^2 \left(\cos 2\theta +3\cos 2\varphi \right) \right] \Bigg\} \, .
\end{align}

\bibliography{biblio_alpha}

\end{document}